\documentstyle[12pt]{article}

\def\be{\begin{equation}}
\def\ee{\end{equation}}
\def\bea{\begin{eqnarray}}
\def\eea{\end{eqnarray}}

\def\br{\begin{thebibliography}{abc}}
\def\er{\end{thebibliography}}

\newcommand{\eqn}[1]{(\ref{#1})}
\def\tr{\mathop{\rm tr}\nolimits}
\def\Tr{\mathop{\rm Tr}\nolimits}

\def\cstars{$C^*$-algebras }
\def\cstar{$C^*$-algebra }

\def\unit{I\!\!I}

\def\wt{\widetilde}

\def\ra{\rightarrow}

\def\iff{\Leftrightarrow}

\def\rt{\rightarrow}

\def\a{\alpha}
\def\b{\beta}

\def\d{\delta}

\def\g{\gamma}
\def\h{\eta}

\def\l{\lambda}

\def\o{\omega}

\def\r{\rho}

\def\x{\xi}

\def\D{\Delta}
\def\F{\Phi}

\def\O{\Omega}

\def\real{\relax{\rm I\kern-.18em R}}
\def\com{\relax\,\hbox{$\inbar\kern-.3em{\rm C}$}}
\def\bc{{\bf C}}
\def\br{{\bf R}}

\def\inbar{\,\vrule height1.5ex width.4pt depth0pt}
\def\IC{\relax\,\hbox{$\inbar\kern-.3em{\rm C}$}}
\def\ID{\relax{\rm I\kern-.18em D}}
\def\IF{\relax{\rm I\kern-.18em F}}
\def\IH{\relax{\rm I\kern-.18em H}}
\def\II{\relax{\rm I\kern-.17em I}}
\def\I1{\relax{\rm 1\kern-.28em l}}
\def\IN{\relax{\rm I\kern-.18em N}}
\def\IP{\relax{\rm I\kern-.18em P}}
\def\IQ{\relax\,\hbox{$\inbar\kern-.3em{\rm Q}$}}
\def\IZ{\relax\,\hbox{$\inbar\kern-.3em{\rm Z}$}}
\def\R{\relax{\rm I\kern-.18em R}}
\font\cmss=cmss10 \font\cmsss=cmss10 at 7pt
\def\Z{\relax\ifmmode\mathchoice
{\hbox{\cmss Z\kern-.4em Z}}{\hbox{\cmss Z\kern-.4em Z}}
{\lower.9pt\hbox{\cmsss Z\kern-.4em Z}}
{\lower1.2pt\hbox{\cmsss Z\kern-.4em Z}}\else{\cmss Z\kern-.4em
Z}\fi}
\def\ca{{\cal A}}

\def\cc{{\cal C}}

\def\ce{{\cal E}}
\def\cf{{\cal F}}

\def\cu{{\cal U}}

\def\iff{\Leftrightarrow}

\def\Hat#1{\rlap{\kern.10em$\widehat{\phantom G}$}#1}
\def\HAt#1{\rlap{\kern.05em$\widehat{\phantom G}$}#1}

\def\czp#1{\rlap{\kern.1em$\widehat{\phantom{G\vrule height.8em}}$}#1{}}
\def\Czp#1{\rlap{\kern.05em$\widehat{\phantom{G\vrule height.8em}}$}#1{}}
\def\bra#1{\left\langle #1\right|}
\def\ket#1{\left| #1\right\rangle}

\def\VEV#1{\left\langle #1\right\rangle}
\newcommand{\sect}[1]{\setcounter{equation}{0}\section{#1}}
\newcommand{\subsect}[1]{\subsection{#1}}

\def\fn{\footnote}
\def\sxn#1{\bigskip\medskip \sect{#1} \smallskip
                                                 }
\def\subsxn#1{\medskip \subsect{#1} \smallskip
                                                }

\begin{document}

\thispagestyle{empty}
\setcounter{page}{0}

\begin{flushright}
hep-lat/9604012\\
ESI (1995) 299\\
SU-4240-621\\
DFTUZ/96/06 \\
DSF-T-15/96 \\
UICHEP-TH/95-12 \\
April 1996, Revised December 1996
\end{flushright}

\vspace{.2cm}
\begin{center}
{\LARGE Lattice Gauge Fields and }\\
\vspace{4mm}
{\LARGE  Noncommutative Geometry}\\
\vspace{5mm}
\end{center}
\begin{center}
{\large A.P.~Balachandran$^{1}$,
		    G. Bimonte$^{2}$
                    G. Landi$^{3,4}$, \\}
\vspace{2mm}
{\large  F. Lizzi$^{4,5}$,
                    P. Teotonio-Sobrinho$^{6}$ }
\fn{Address after 03/01/96: Universidade de Sao Paulo, Instituto de Fisica -
DFMA, Caixa Postal 663118, 05389-970 Sao Paulo, SP, Brazil.  }\\
\vspace{.5cm}
{\it The E. Schr\"odinger International Institute for
Mathematical Physics,}\\
{\it Pasteurgasse 6/7, A-1090 Wien, Austria.}\\
\vspace{2mm}
{\it $^1$ Department of Physics, Syracuse University,
Syracuse, NY 13244-1130, USA.}\\
\vspace{2mm}
{\it $^2$ Departamento de Fisica Teorica, Facultad de
Ciencias,}\\
{\it Universitad de Zaragoza, 50009 Zaragoza, Spain.}\\
\vspace{2mm}
{\it $^3$ Dipartimento di Scienze Matematiche,
Universit\`a di Trieste,}\\
{\it P.le Europa 1, I-34127, Trieste, Italy.}\\
\vspace{2mm}
{\it $^4$ INFN, Sezione di Napoli, Napoli, Italy.}\\
\vspace{2mm}
{\it $^5$ Dipartimento di Scienze Fisiche, Universit\`a di
Napoli,}\\
{\it Mostra d' Oltremare, Pad. 19, I-80125, Napoli, Italy.}\\
\vspace{2mm}
{ \it $^6$ Department of Physics, University of  Illinois at Chicago,}\\
{\it Chicago, IL 60607-7059, USA.}\\
\end{center}
\vspace{.2cm}

\vfill\eject
\begin{abstract}
Conventional approaches to lattice gauge theories do not
properly consider the topology of spacetime or of its fields. In this
paper, we develop a formulation which tries to remedy this defect. It starts
from a cubical decomposition of the supporting manifold
(compactified spacetime or spatial slice) interpreting it as a finite 
topological
approximation in the sense of Sorkin. This finite space is entirely described
by the algebra of cochains with the cup product. The methods of Connes and
Lott are then used to develop gauge theories on this algebra and to derive
Wilson's actions for the gauge and Dirac fields therefrom which can now be given
geometrical meaning. We also describe very natural candidates for the QCD
$\theta
$-term and Chern-Simons action suggested by  this algebraic formulation.
Some of these formulations are simpler than currently available
alternatives. The paper treats both the functional integral and Hamiltonian
approaches.

 Geometry and topology of complexes. 
Noncommutative Geometry. Topological Actions. Chern-Simons terms.



\end{abstract}

\newpage

\sxn{Introduction}

There is general consensus that elementary particle interactions below the
scale of 1 Tev are accurately described by the standard model \cite{GSW}.
It is a
gauge theory of strong and weak interactions based on the hypothesis that
spacetime is a manifold. Strong processes and weak phenomenology give
persuasive evidence that it is a good model at these energies, at least
whenever perturbation theory gives a decent reliable approximation.

Nevertheless several central issues in elementary particle 
physics cannot
be addressed using perturbation theory. Examples are problems concerning
the bound state spectrum, quark confinement and chiral symmetry breaking in
QCD. Numerical schemes, known generically as lattice gauge theories
\cite{LGT}, have been developed during the past several years because of
limitations of perturbation theory and other analytical methods, and much
effort has also been spent studying QCD with their help.

In previous work \cite{BBET,lisbon,ncl}, we have emphasized that
conventional formulations of  lattice gauge theories are naive in their
treatment of topology. They typically begin from an approximation of the
supporting manifold (spacetime or spatial slice) with a finite set of
points having the trivial discrete topology (each point being both open
and closed), and with no further structure. This rough and crude 
attitude precludes transparent and
convincing representations of continuum topology. In fact, it even
provokes the suspicion that lattice models are incapable of describing the
continuum in a good manner at any finite level of approximation.

This paper attempts to remedy this defect. It starts with the observation
that cell decompositions of the manifold, such as cubical and simplicial
ones, are associated with certain open covers which in our context can
also be realistically assumed to be finite. They are thus finite
topological spaces, being examples of partially ordered sets or posets of
Sorkin \cite{SORK,related,TEOTONIO}. The latter have been treated in detail
elsewhere and shown capable of reproducing subtle continuum features
significant for quantum physics \cite{SORK,BBET,ncl}.

Now the chain complex generated by cells are ``dual'' to an associative
non-commutative differential algebra $\Omega ^*{\cal A}$ which encodes all
the information in the chains and can reproduce the latter in full detail.
The algebra $\Omega ^*{\cal A}$ is just a variant of the algebra of cochains
under the cup product \cite{MICHOR}
\footnote{ A previous approach using a
nonassociative modification of the algebra of cochains for lattice gauge theories
is described in \cite{SSZ}. {Nonassociative} 
algebras like those in \cite{SSZ} have also been used in \cite{BR} to 
develop discrete
topological field theories.}. It is even possible to reproduce the poset
from its differential ideals just as a Hausdorff space can be reproduced
from its \cstar of continuous functions  using the Gel'fand-Naimark 
construction \cite{FD}. In this
manner, we encode the topology of the approximating set in an algebra.

There are distinct advantages to working with $\Omega ^*\ca$ instead of the
chains because of the beautiful methods from Connes, Lott and others
\cite{CONLOT,CONNESBOOK,VARI}
for writing gauge theories starting from algebras. By
adapting them to $\Omega ^*\ca$, we are able to show that exactly the
Wilson actions \cite{LGT} for the glue and Fermi fields are reproduced by
these methods for a cubical chain complex. These actions thus naturally
emerge from an approach to discretisation which pays conscious attention
to the continuum topology.

The action of QCD has a topological term, namely the integral of 
${\theta \over 8\pi^2} \Tr F\wedge F$,  with $F$ being the curvature. This 
``$\theta $-term'' ,
or topological susceptibility  is `topological' since it is independent  of
spacetime metric. The ``strong CP problem'' is due to its existence. Its 
integrand also governs the chiral anomaly and has a central role in studies of
chiral symmetry breakdown, the
$U(1)$  problem and decays like $\pi ^0\rt 2\gamma $. There are thus numerous
reasons for finding a lattice analogue of the $\theta $-term. Several lattice
versions have also been proposed
\cite{Luscher,LGT,GKLSW,Teper}, but none is truly natural. The algebraic
approach too has a suggestion for this term. It is very natural, being unique,
and involves fewer Wilson links than existing proposals. 
In addition, the algebraic approach has natural discrete versions for the
topological term of two-dimensional QED and also the Chern-Simons \cite{CS}
term.  [For previous discretisations of the latter see
\cite{PhillipsStone,Semenoff}.]

In what follows, we will be discussing these topological terms in the
context of the algebraic approach. Exploration of their properties by
lattice theorists is indicated, as they may be viable 
substitutes to existing versions.

The paper is organized as follows. Section 2 is a short introduction to
Sorkin's posets \cite{SORK} and their relation to chain decompositions of
manifolds. Section 3 discusses the cochain algebra while Section 4
formulates integration theory using this algebra. Sections 5 to 10 limit
themselves to $\Omega^*\ca$ for cubical decompositions. Section 5
introduces the algebra $\Omega^*\ca$ and applies the ideas of Connes and Lott
\cite{CONLOT,CONNESBOOK,VARI} to show the emergence of the Wilson action in a
simple way from the algebraic approach. Section 6 does the same to derive 
Wilson's fermion action from the algebraic approach. Section 7 studies the 2d
topological action while Sections 8 and 9  extend this discussion to the four
dimensional case, and to CS terms in one and three dimensions. The material in
Sections 5 to 9 is in the context of functional integrals, as they deal with
actions to be fed into these integrals. In Section 10, we also sketch how a
Hamiltonian formulation can be derived taking advantage of an earlier work of
Rajeev \cite{Rajeev}. The final Section 11 outlines possible generalizations of
the algebraic approach to simplicial and other discretisations of the supporting
manifold. They can be relevant because cubical decompositions are
convenient only for toroidal topologies, and so for spacetime or spatial
manifolds compactified with periodic boundary conditions. 
In addition this section speculates about the
reason why the QCD $\theta$ is small and also points out the 
advantages of the algebraic approach in
subjects like soliton physics, which are not gauge theories.

In previous work \cite{lisbon,ncl,PANG}, we had associated an algebra $\ca$,
different from $\Omega^*\ca$, with Sorkin's topological approximations.
That algebra was $C^*$ unlike $\Omega^*\ca$. It 
was also infinite
-dimensional creating problems of interpretation and adaptation to
numerical use. Although not a \cstar, $\Omega ^*\ca$ 
does not have these
problems and therefore may be superior to  $\ca$.

\sxn{Noncommutative Lattices and Chains}

In this Section we describe the topological approximation method for
topological spaces developed previously by Sorkin \cite{SORK}. We will show that
a finite (or finitary) approximation to a topological space can be described by 
a  poset. This poset with its topology is the structure space of a noncommutative
algebra. Hence the name {\it Noncommutative Lattices}, which we use
interchangeably with the word poset.

\subsxn{Posets}

The word poset \cite{Alex} is an acronym for a ``partially ordered set''.
We will encounter
only finite posets, so we assume that all our posets are finite.
Generalization to the countable case is straightforward.
If $P= \{p_1,p_2,...,p_K\}$ is a poset with $K$ elements, then by
definition,
there is an order relation $\preceq $ between some pairs of 
its points such that
{\em i}) $p_i\preceq p_j$ and $p_j\preceq p_k$ $\Rightarrow $ $p_i\preceq p_k$
(transitivity) and {\em ii}) $p_i\preceq p_i$.

A poset has a canonical topology derived from its partial order. A basis of
open sets $O_j$ for this topology is as follows : the smallest open set $O_j$
containing $p_j$ consists of all $p_k$ such that $p_k\preceq p_j$:
\be
O_j=\{p_k~:~p_k\preceq p_j\}.\label{2.1}
\ee

Sorkin \cite{SORK} has shown that every finite open cover of a manifold
$M$ leads to a (finite) poset. Also a simplicial or cubical decomposition
leads to a poset. This result will be illustrated below. 
It has also been shown \cite{SORK,PANG}
that $M$ can be recovered as a topological space from a repeated
refinement of the covers and a suitable inverse limit.

\subsxn{The Circle and its Noncommutative Lattice}

Let us work out  the example of the 
circle in some detail.
The circle $S^1$ can be discretised using the lattice $L_N$ with $N$
points having coordinates $z_k=e^{ik\frac{2\pi }{N}}$, $k\in \mbox{\Z}$
mod $N$.
This lattice gives rise to a cubical decomposition of $S^1$ with zero and one
cells
\be
C_0=\{z_0,z_1,...,z_{N-1}\};~~~
C_1=\{z_{0,1},z_{1,2},...,z_{N-1,N}\}\label{2.2}
\ee
where $z_{j,j+1}:= \mbox{ the interval }(z_j,z_{j+1})$.
The elements of the vector space over $\mbox{\IC}$ freely generated by these
cells
are the chains. We can also introduce a boundary operator $\partial $ on these
chains according to
\be
\partial z_j=0,~~~\partial z_{j,j+1}=z_{j+1}-z_j\ \ , \label{2.3}
\ee
$\partial^2$ being consequently  zero. Here $z_j$ and $z_{j+1}$ are to be
regarded as the boundaries of $z_{j,j+1}$. The chains become a chain
complex in the presence of $\partial $.

There is a natural noncommutative lattice
 associated with $C_0$ and $C_1$. We set
$z_j\preceq z_j$, $z_{j,j+1}\preceq z_{j,j+1}$ and $z_{k,k+1}\preceq z_j$ if
$j\in \{k,k+1\}$. The resultant noncommutative lattice 
can be conveniently shown with the
aid of a Hasse diagram \cite{Alex} as in Fig. 1.
In this diagram, if two points $a$ and $b$ are connected by a link,
and $b$ is lower than $a$, then $b\preceq a$.
\begin{figure}[ht]
\begin{center}
\unitlength=1.00mm
\linethickness{0.4pt}
\begin{picture}(66.00,39.00)
\put(5.00,15.00){\circle*{2.00}}
\put(20.00,15.00){\circle*{2.00}}
\put(35.00,15.00){\circle*{2.00}}
\put(5.00,35.00){\circle*{2.00}}
\put(20.00,35.00){\circle*{2.00}}
\put(35.00,35.00){\circle*{2.00}}
\put(5.00,35.00){\line(0,-1){20.00}}
\put(5.00,15.00){\line(3,4){15.00}}
\put(20.00,35.00){\line(0,-1){20.00}}
\put(20.00,15.00){\line(3,4){15.00}}
\put(35.00,35.00){\line(0,-1){20.00}}
\put(65.00,15.00){\circle*{2.00}}
\put(65.00,35.00){\circle*{2.00}}
\put(65.00,35.00){\circle*{2.00}}
\put(65.00,35.00){\line(0,-1){20.00}}
\put(47.00,15.00){\makebox(0,0)[cc]{. . .}}
\put(47.00,35.00){\makebox(0,0)[cc]{. . .}}
\put(35.00,15.00){\line(3,4){10.67}}
\put(5.00,35.00){\line(3,-1){60.00}}
\put(65.00,35.00){\line(-5,-6){9.00}}
\put(5.00,11.00){\makebox(0,0)[cc]{$z_{0,1}$}}
\put(20.00,11.00){\makebox(0,0)[cc]{$z_{1,2}$}}
\put(65.00,11.00){\makebox(0,0)[cc]{$z_{N-1,N}$}}
\put(65.00,39.00){\makebox(0,0)[cc]{$z_{N-1}$}}
\put(35.00,39.00){\makebox(0,0)[cc]{$z_{2}$}}
\put(20.00,39.00){\makebox(0,0)[cc]{$z_1$}}
\put(5.00,39.00){\makebox(0,0)[cc]{$z_0=z_N$}}
\end{picture}
\end{center}
\begin{center}
{\footnotesize {\bf Fig. 1.} The Hasse diagram of the circle poset.}
\end{center}
\end{figure}

In Hasse diagrams with many levels, the corresponding statement is as
follows. If $\xi _0$ and $\xi _L$ are two points in the diagram and it has
points $\xi _1, \xi _2, ..., \xi _{L-1}$ such that $\xi _i$ and $\xi _{i+1}$
are linked  and $\xi _i$ is lower than $\xi _{i+1}$, then
$\xi _0\preceq \xi _L$. See Fig. 2 for the example of a two
sphere poset taken
from Sorkin \cite{SORK}. Note that the partial order rule of the two--level
diagram plus transitivity gives the rule for any diagram.
\begin{figure}[ht]
\begin{center}
\unitlength=1.00mm
\linethickness{0.4pt}
\begin{picture}(35.00,46.00)
\put(10.00,15.00){\circle*{2.00}}
\put(10.00,30.00){\circle*{2.00}}
\put(10.00,45.00){\circle*{2.00}}
\put(30.00,15.00){\circle*{2.00}}
\put(30.00,30.00){\circle*{2.00}}
\put(30.00,45.00){\circle*{2.00}}
\put(30.00,45.00){\line(0,-1){30.00}}
\put(10.00,15.00){\line(0,1){30.00}}
\put(10.00,45.00){\line(4,-3){20.00}}
\put(30.00,30.00){\line(-4,-3){20.00}}
\put(30.00,45.00){\line(-4,-3){20.00}}
\put(10.00,30.00){\line(4,-3){20.00}}
\put(35.00,15.00){\makebox(0,0)[cc]{$b_1$}}
\put(35.00,30.00){\makebox(0,0)[cc]{$b_2$}}
\put(35.00,45.00){\makebox(0,0)[cc]{$b_3$}}
\put(5.00,45.00){\makebox(0,0)[cc]{$a_3$}}
\put(5.00,30.00){\makebox(0,0)[cc]{$a_2$}}
\put(5.00,15.00){\makebox(0,0)[cc]{$a_1$}}
\end{picture}
\end{center}
\begin{center}
{\footnotesize {\bf Fig. 2.} A two-sphere poset. Here $a_j\preceq a_k$ or
$b_k$, $b_j\preceq a_k$ or $b_k$ whenever $j<k$.}
\end{center}
\end{figure}

According to poset theory, the basis of open sets for the poset $P$ of Fig. 1
consists of the following:
\be
O_{j, j+1}=\{z_{j,j+1}\}, ~~~~~O_j=\{z_j, z_{j-1,j}, z_{j,j+1}\}.\label{2.4}
\ee
What is the relation of these open sets to those of $S^1$ ?

There is in fact a natural correspondence between these open sets. Let us
assume that $z_{j,j+1}$, represents the open interval $\{e^{i\theta
\frac{2\pi }{N}}: j<\theta <j+1\}$ of $S^1$. Then one sees right away that
$O_j$ and $O_{j,j+1}$ are open in $S^1$. More pedantically , we can
construct a map $\varphi :S^1\rt P$ by setting $\varphi (p)=p$ if $p=z_i$
and $\varphi (p)=z_{j,j+1}$ if $p\in z_{j,j+1}$. Then $\varphi ^{-1}(O_j)$
and $\varphi ^{-1}(O_{j,j+1})$ are open or $\varphi $ is continuous. The
topology of $P$ is in fact the quotient topology induced by the map
$\varphi $.

One further point may be made. The open sets $\varphi ^{-1}(O_j)$ and
$\varphi ^{-1}(O_{j,j+1})$ provide an
open cover of $S^1$. Suppose now that we identify any two points if
neither or
both are in every one of these open sets. Then we get back $P$, $\varphi $
being just this identification map. This construction of $P$ from an open
cover is an example of Sorkin's construction \cite{SORK} of a poset from a
manifold $M$.

The interpretation of $P$ as a topological approximation of $M$ has been
successfully made by Sorkin \cite{SORK}. The physical meaning and mathematical
power of this approximation have been dealt with elsewhere
\cite{SORK,PANG,BBET,lisbon,ncl,SorkinGrav}.
It is in this way that topological data of $M$ are stored in its chains.

These considerations easily extend to any chain complex, such as a simplicial
complex (obtained from a simplicial decomposition) of a generic manifold $M$.
Thus supposing
that $\a_{(n)}^l$ denotes an $n-$cell, the partial order is introduced by
setting $\a_{(m+1)}^k \preceq \a_{(m)}^j$ if $\a^j_{(m)}$ is a face of
$\a_{(m+1)}^k$. The  resultant poset $P$ is a topological space.

If now we regard each cell as consisting only of its interior points, then
the open sets of $P$ lead as before to an open cover of $M$. The
identification map $\varphi $ from $M$ to $P$ can also be constructed as
we did above, the poset topology being just the quotient topology induced by
$\varphi $.

Thus always the cells of a manifold $M$ and their incidence relations
\cite{Hom} define a finite topological space $P$ approximating $M$ including its
topology.

\sxn{The Cup Algebra}

In this section we define a differential calculus for forms and 
chains on noncommutative lattices, based on the cup product. 
Differential calculi on noncommutative spaces have been discussed 
elsewhere \cite{diffcal} in different contexts.

\subsxn{Construction}

In the usual way \cite{Hom}, we associate a complex vector space of chains with
the preceding cells by taking their complex linear combinations. We also define
a boundary operator $\partial $ such that $\partial ^2=0$:
\be
\partial \a_{(m)}^j= \sum _k I\left( \a_{(m)}^j, \a_{(m-1)}^k \right)
\a_{(m-1)}^k,~~~I\left( \a_{(m)}^j, \a_{(m-1)}^k \right)=\pm 1\mbox{ or }
0~,\label{3.1}
\ee
\be
\partial ^2=0~~\mbox{ or }~~
\sum _k I\left( \a_{(m)}^j, \a_{(m-1)}^k \right)
		I\left( \a_{(m-1)}^k, \a_{(m-2)}^l \right)=0~.
\ee
Here $I\left( \a_{(m)}^j, \a_{(m-1)}^k \right)$ are incidence numbers. The
chain complex we have here will be called $\cc$.

Let $\cc^*$ be the vector space dual to $\cc$. It has a basis ${\a^{(n)}_j}$
defined by
\be
\VEV{\a^{(m)}_j, \a_{(n)}^k }=\delta^m{}_n \delta_j{}^k~,
\label{3.3}
\ee
$\VEV{\cdot,\cdot}$ being the duality pairing.

The dual $d$ of $\partial $ is given by the relation 
\be
\VEV{ d{\a^{(m)}_j}, \a_{(n)}^k }=
\VEV{ {\a_j{}^{(m)}}, \partial \a_{(n)}^k }.\label{3.4}
\ee
It is nilpotent:
\be
d^2=0.\label{3.5}
\ee

The cochain complex $\cc^*$ can be made into an algebra by defining a
product of its elements. This product is the cup product \cite{MICHOR}
which is the discrete analogue of the wedge product on differential forms.
It can be introduced as follows. One first orders the vertices of $\cc$ by
numbering them from $0$ to $N-1$. Any oriented cell is given by its ordered
vertices, an odd permutation of the order changing its sign. It can
therefore be written as $[i_0,i_1,...,i_k]=(-1)^P~[j_0,j_1,...,j_k]$ where
$j_0<j_1<...<j_k$ and $(-1)^P$ is the sign of the permutation $P$ bringing
the entries on the left to those on the right. We can then label the
elements of the dual basis by $[j_0,j_1,...,j_k]^*=(-1)^P~[i_0,i_1,...,i_k]^*$,
$[j_0,j_1,...,j_k]^*$ giving 1 on pairing with $[j_0,j_1,...,j_k]$ and zero
with a cell having any vertex different.

With this convention at hand, the cup product $\sqcup $  is defined as
follows:
$$
[a_0,a_1,...,a_k]^*\sqcup [b_0,b_1,...,b_l]^*
$$
is a cochain nonvanishing only on a cell with $k+l+1$ vertices. So we must give
its value on $[c_0,c_1,...,c_{k+l}]$. We first write the last as
$\epsilon[d_0,d_1,...,d_{k+l}]$ where $\epsilon$ is 1 or -1 and
$d_0<d_1<...<d_{k+l}$. Then
$$
\VEV{[a_0,a_1,...,a_k]^*\sqcup [b_0,b_1,...,b_l]^*, [c_0,c_1,...,c_{k+l}]}=
$$
\be
=\epsilon\VEV{[a_0,a_1,...,a_k]^*,[d_0,d_1,...,d_{k}]}~
\VEV{[b_0,b_1,...,b_l]^*,[d_k,d_{k+1},...,d_{k+l}]}.\label{3.6}
\ee

The cup product is associative.

We call the (associative) algebra of cochains under the cup product as
$\omega ^*\ca$, reserving the symbol $\Omega ^*\ca$ for its modifications,
to be introduced later. The ordering of vertices, and hence $\sqcup $ not
being unique, $\omega ^*\ca$ is also not unique. It does become unique when
projected down to cohomology classes.

The algebra $\ca$, wherefrom the notation $\o^*\ca$ is derived,
consists of zero cochains and is abelian. We also denote it by
$\o^{(0)}\ca$, $\o^{(k)}\ca$ signifying $k-$cochains, non-vanishing
only on chains with $(k+1)$ vertices. Thus $\o^*\ca$ has a natural grading,
being $\oplus_k \o^{(k)}\ca$, with $(\o^{(k)}\ca) (\o^{(l)}\ca) \subseteq
\o^{(k+l)}\ca$. The operator $d$ is of grade or degree $1$ since
$d\o^{(k)}\ca \subseteq \o^{(k+1)}\ca$.

The elements of $\o^{(k)}\ca$ can be legitimately regarded as
approximations to differential $k$-forms, and $\o^*\ca$ as
approximating the algebra of differential forms. But there is one
notable difference: in general
\be
 \a^{(k)}  \a^{(l)} \neq (-1)^{kl} \a^{(l)} \a^{(k)}~, ~~ \a^{(k)} \in
\o^{(k)}\ca~, \a^{(l)} \in \o^{(l)}\ca~. \label{3.7}
\ee
In other words, \underline{$\o^*\ca$ is not graded--commutative} in contrast to
the algebra of differential forms. This fact will become a source of
unpleasantness when formulating gauge theories.

In analogy to usage for manifolds, we shall call the elements of
$\o^{(k)}\ca$ as $k$-forms. 

\subsxn{Examples}
Let us look at cubical decomposition of tori as examples.

We begin with $S^1$ and the lattice $L_N$. The algebra $\ca$ then consists of
functions on $L_N$ with the basis $\{z^j\}$ defined by
\be
\VEV{z^j,z_k}=\delta^j{}_k\ \ .
\ee
The one-forms give the $\ca$-module generated
by the duals $z^{i,i+1}$ of $z_{i,i+1}$ defined by
$\VEV{z^{i,i+1},z_{j,j+1}}= \delta^i{}_j$. There are no forms of larger
order. 
Of course the definition of
the $\ca$-module involves the cup product. For the order given
by the subscripts on $z_j$'s, one verifies it to be 
\bea
z^j  \sqcup z^k  & = & \d_{j k} z^j~, \nonumber \\
z^k \sqcup z^{j, j+1} & = & \d_{jk} z^{j, j +1}~, \nonumber \\
z^{j, j+1}  \sqcup z^k  & = & \d_{j+1, k} z^{j, j+1}~, \nonumber \\
z^{j, j+1} \sqcup z^{k, k+1} & = & 0~,
\label{3.8}
\eea
where, as would be expected, indices differing by $N$ are to be identified.

These product rules can be written in terms of components of elements. Two
general elements of the algebra can be written as
\bea
\a &=& \sum \a_i z^i + \sum \a_{i, i+1} z^{i, i+1}~, \nonumber\\
\b &=& \sum \b_k z^k + \sum \b_{k, k+1} z^{k, k+1}~, \label{3.9}
\eea
so that coordinate representations of $\a$ and $\b$ are 
\bea
\a &=& (\a_{0}, \a_{1}, \dots, \a_{N-1}; \a_{0,1}, \a_{1,2}, 
\dots \a_{N-1, N})~,
\nonumber\\
\b &=& (\b_{0}, \b_{1}, \dots, \b_{N-1}; \b_{0,1}, \b_{1,2}, 
\dots \b_{N-1, N})~,
\label{3.10}
\eea
while the cup product can also be written as
\bea
\a \b &=& (\a_0 \b_0, \a_1 \b_1, \cdots , \a_{N-1} \b_{N-1} ;
\nonumber \\
&&\a_0 \b_{0,1} + \a_{0,1} \b_1,
\a_1 \b_{1,2} + \a_{1,2} \b_2, \cdots,
       \a_{N-1} \b_{N-1,N} + \a_{N-1,N} \b_N)~. \label{3.11}
\eea

The (left) regular representation of this algebra in the basis $\{z^{01},
z^{12},\ldots,z^{N-1,N}; z^0,z^1,\ldots,z^{N-1}\}$ is 
\bea
 \sum \a_j z^j &\ra&
\mbox{diag}[\a_{0}, \a_{1}, \dots,  \a_{N-1}; \a_{0}, \a_{1}, \dots, 
\a_{N-1}]~,\nonumber \\
 \sum \a_{j, j+1} z^{j ,j+1} &\ra&
\left[
\begin{array}{cc}
\mbox{\huge 0} &
\begin{array}{cccccccc}
0 & \a_{0,1} & 0 & \cdot & \cdot & \cdot & 0 \\
0 & 0 & \a_{1,2} & 0 & ~ & ~ & 0 \\
\cdot & \cdot & 0 & \cdot & ~ & ~ & ~ \cdot \\
0 & 0 & \cdot & \cdot & \cdot & 0 & \a_{N-2,N-1} \\
\a_{N-1,0} & 0 & \cdot & \cdot & \cdot & 0 & 0
\end{array}\\
\mbox{\huge 0} & \mbox{\huge 0}
\end{array}
\right]~.
\label{3.12}
\eea  

Now, $\o^* \ca$ is a differential algebra with a differential $d$ which
is the dual of the following boundary operator $\partial$ on chains:
\bea
\partial z_j &=& 0~,\nonumber\\
\partial z_{j,j+1} &=& z_{j+1}-z_j\, ,\nonumber\\
\partial^2 &=& 0~.
\label{3.13}
\eea
It is given by 
\bea
d z^j &=& z^{j-1,j}-z^{j,j+1}~, \nonumber\\
d z^{j, j+1} &=& 0~, \label{3.14}
\eea
so that $d^2 = 0$.

It turns out that for cubical decompositions, $d$ can be realized in terms of a
linear operator $F_+$ in the regular representation, much as in cyclic
cohomology \cite{CONNESBOOK}, the relevant formula being
\be
d \a^{(k)} = F_+ \a^{(k)} - (-1)^k \a^{(k)} F_+~, \label{3.15}
\ee
where we have identified $\o^*\ca$ with its representation. In fact $F_+ \in
\o^{(1)}\ca$. For the present case of $S^1$, the formula for $F_+$ is
\be
F_+ = \left(
\begin{array}{cc}
0 & \D \\
0 & 0
\end{array}
\right) ~, ~~ \D_{rs} = \d_{r,s-1}~.     \label{3.16}
\ee
It may be remarked that 
\be
\D\left(\begin{array}{ccc}
\a_0 & &\\
& \ddots &\\
& & \a_{N-1}
\end{array}\right) \D^\dagger=
\left(\begin{array}{ccc}
\a_1 & &\\
& \ddots &\\
& & \a_{N}
\end{array}\right) \ ;\ \ \a_N=\a_0 , \label{rotaz}
\ee
that is,  it consists of $\a_i$ rotated by
$e^{+i 2\pi / N}$. We have the identities
\bea
\D^{\dag} &=& \D^T = \D^{-1}~, \nonumber\\
\D^N &=& \unit ~, ~\mbox{det} \D = (-1)^{N-1}~. \label{3.17}
\eea

Later on, we will modify this algebra to another one, more appropriate for
gauge theories, to work with.

Note that
\be
F = \left(
\begin{array}{cc}
0 & \D \\
\D^{\dag} & 0
\end{array}
\right) ~,
~\wt{F} = i \left(
\begin{array}{cc}
0 & - \D \\
\D^{\dag} & 0
\end{array}
\right) ~~~~ {\rm and}~~
\g = \left(
\begin{array}{cc}
\unit  & 0 \\
0 & -\unit
\end{array}
\right) 
\label{3.18}
\ee
generate a Clifford algebra (isomorphic to that of Pauli matrices) and
that $F_+ = {1 \over 2} (F + i \wt{F})$ anti-commutes with $\g$. Furthermore
\be
F_+^2=0~.
\ee

Generalization to higher dimensions is easy. Consider for example $S^1 \times
S^1 := T^2$. We can assume that its cubical cells are given by the Cartesian
product $L_N \times L_N$. As for its differential algebra, first let
\be
\o^*\ca ^{(1)} := \o^*\ca \otimes_{\bc} \unit~, \label {3.19}
\ee
$\o^*\ca$ being for $L_N$ and $\unit$ being the constant function with value
$1$. The algebra $\o^*\ca^{(1)}$ will consist of elements describing the
first $L_N$ in the product $L_N \times L_N$. If $\o^*\ca ^{(2)}$ is to be
associated with the second $L_N$, its one-forms must anti-commute with
those of $\o^*\ca ^{(1)}$. For this reason, set
\be
\o^*\ca ^{(2)} = \unit \otimes \ca \oplus \g \otimes \o^{(1)}\ca :=
\o^{(0)}\ca^{(2)} \oplus \o^{(1)}\ca ^{(2)}
:= \ca ^{(2)} \oplus \o^{(1)}\ca ^{(2)}~. \label{3.20}
\ee
Then the algebra of $L_N \times L_N$, which we again call $\o^*\ca$, is
generated by $\o^*\ca ^{(1)}$ and $\o^*\ca ^{(2)}$,
\be
\o^*\ca  = \o^*\ca ^{(1)} \cdot \o^*\ca ^{(2)}~. \label{3.21}
\ee

The differential $d$ for this algebra is also readily found. Thus let
\be
F_+^{(1)} = f_+ \otimes \unit~, ~~~ F_+^{(2)} = \g \otimes f_+~,
\label{3.22}
\ee
where $f_+$ is the same matrix which appears in (\ref{3.16}). [The notation
has been slightly changed to avoid confusion below.] Then $d$ is defined by
(\ref{3.15}) where
\be
F_+ = F_+^{(1)} + F_+^{(2)}~, ~~~ F_+F_+ \equiv F_+^2 = 0~. \label{3.23}
\ee

\subsxn{The Reconstruction Theorem}\label{se:rt}

There is a nice way to reconstruct the poset $P$ from the algebra
$\o^*\ca$. It very much resembles the methods used in topologising the
structure space (or space of irreducible $^*$-representations (IRR's)) of
\cstars \cite{FD} when the former for example is a finite set.
We can associate ideals [``primitive ideals''] with points $p$ of
the structure space, the ideal $I_p$ at $p$ being the kernel of the IRR
$p$ [elements which vanish at $p$]. There is a distinct $I_p$ at distinct
$p$. We can now partially order $I_p$ by inclusion: $I_p \preceq I_q \iff I_p
\subseteq I_q$. This gives a partial order also on $p$'s: $p \preceq q \iff
I_p \preceq I_q$. In this way, the structure space becomes a poset with its
topology. The latter is the analogue of the Gel'fand topology for the structure
space of abelian \cstars and equivalent to the hull-kernel topology
\cite{FD}. As emphasized elsewhere \cite{BBET,lisbon,BBMS}, this
manner of retrieving topology from algebras seems to be of exceptional
importance for foundations of quantum physics.

Now, it is the case that $\o^*\ca$ is not a \cstar, as it does not
contain the adjoint of forms of nonvanishing degree. So we must
suitably modify the rules of reconstruction of $P$ from its ideals.
For this purpose, let us define a {\em differential ideal} of
$\o^*\ca$ as a two-sided ideal closed 
under $d$. Let us then say that a
differential ideal $I$ is {\em indecomposable} if it is not the
intersection of two other differential ideals $I_i$ both distinct from $I$ 
$: I \not= I_1 \cap I_2$, for $I_1\neq I_2$ and $I\not= I_1$ or $I_2$.

If two representations have ideals $I_1$ and $I_2$, their direct
sum has the ideal
$I_1\cap I_2$. Thus the notion of indecomposability here replaces the
notion of irreducibility. The latter is not appropriate for us. Our
algebra has nilpotent elements and has incompletely reducible faithful
representations which we will have to consider.

Let $\{I_\a \}$ be the set of indecomposable differential ideals of $\o^*\ca$.
They can be partially ordered by inclusion just as in the case above:
\be
I_p \preceq I_q \iff I_p \subseteq I_q~. \label{3.24}
\ee
With this partial order, $\{I_\a \}$ becomes a poset which
is just the poset $P$.

We content ourselves by verifying this assertion for the circle case.
For the latter, the indecomposable differential ideals are 
\bea
&& I_j = \{ \sum\a_k z^k + \sum\a_{k,k+1} z^{k,k+1}~:~ \a_j = 0 \}~,
\nonumber \\
&& I_{j,j+1} = \{ \sum\a_k z^k + \sum\a_{k,k+1} z^{k,k+1}~:~ \a_j =
\a_{j, j+1} = \a_{j+1} = 0 \}~. \label{3.25}
\eea                                                       
The proof is by straightforward calculation. Now $I_{j,j+1} \preceq I_j$
and $I_{j,j+1} \preceq I_{j+1}$ by the previous rule, so that the poset
of these ideals is exactly that of Fig. 1.

\sxn{Integration Theory}

A cochain can be evaluated or ``integrated" on a chain. We want to express
these integrals now using Hilbert spaces traces (partial and otherwise) as in
noncommutative geometry \cite{CONNESBOOK}-\cite{VARI},
this formulation being very useful for developing gauge theories.

Let us consider the $S^1$ case first. Here the pairing of a cochain
$\a = \sum \a_i z^i + \sum \a_{i, i+1} z^{i, i+1}$ with a chain
$\x = \sum \x^j z_j + \sum \x^{j, j+1} z_{j, j+1}$ is  
\be\label{4.1}
\VEV{\a, \x} = \sum \a_j \x^j + \sum \a_{j, j+1} \x^{j, j+1}~.
\ee                                                        

We can express this by using matrix elements of operators by setting 
\bea
\d_{jk} &=& \VEV{z^j, z_k} : =  \int_{z_k} z^j \equiv \VEV{ k | z^j | k }
= \VEV{ k+N | z^j | k+N }~, \label{4.2} \\
\d_{jk} &=& \VEV{z^{j, j+1}, z_{k, k+1}} :=  \int_{z_{k, k+1}} z^{j,j+1}
\equiv \VEV{ k | z^{j,j+1} F_+^ \dagger | k} ~, \label{4.3}
\eea
the pairing of zero- (one-) forms with one- (zero-) chains being
identified with zero. Here and in what follows, $z^j$ and
$z^{j,j+1}$ appearing in matrix elements and traces are the
operators of (\ref{3.12}), while $|k>~ \equiv |k + N>~ (0 \leq k \leq N-1)$ is
the column vector with the row corresponding to the site $e^{i k 2 \pi / N}$
equal to $1$, and the remaining rows zero. Its lower half is in particular
zero. From (\ref{4.2},\ref{4.3}), we get the
general formula for (\ref{4.1}) using linearity. Note that the
integral of a one-form $\o$ over the entire circle poset is
\be
\int \o = \Tr\ \o F_+^\dagger  = \sum_{i=1}^N \VEV{i | \o F_+^\dagger | i}~, 
\label{4.4}
\ee
while the partial integral from $j$ to $j+s+1$ is 
\be
\int_{j}^{j+s+1} \o = \sum_{i=j}^{j+s} \VEV{i | \o F_+^\dagger | i}~.
\label{4.5}
\ee                                    
If $\o$ is exact, that is if $\o = d\a = [F_+, \a]$, $\a=\sum\a_iz^i$
being a function, then  
\be
\int_{j}^{j+s+1} \o = \a_{j+s+1} - \a_{j}~,
\label{4.6}
\ee                           
so that Stokes' theorem is valid. The manipulations leading to (\ref{4.6}), are
indicated by                  
\bea
\sum_{i=j}^{j+s}\VEV{i|[F_+,\a]F_+^\dagger |i}&=&\sum_{i=j}^{j+s}  
\left[ 
\D  
\left(
\begin{array}{cccc}  
\a_0 & & &\\
& \a_1 & &\\
& & \ddots & \\
& & & \a_{N-1}
\end{array}
\right)
\D^\dagger-
\left(
\begin{array}{cccc}
\a_0 & & &\\
& \a_1 & &\\
& & \ddots & \\
& & & \a_{N-1}
\end{array}
\right)
\right]_{ii}\nonumber\\
&=&
\a_{j+s+1}-\a_j~.
\label{4.7}
\eea 

It is noteworthy that the traces choose an orientation to the
integrals, as we could equally well have expected to get the negatives
of the above 
right-hand sides. It is in fact possible to get the
reversed sign with a different cup product. A formalism incorporating
both signs can also be developed as we shall later see.

Next consider the two-torus. Its lattice $L_N \times L_N$ can be
labeled by $(\a, \b), ~\a, \b = 0, \dots, N-1. $ On this lattice, we
can integrate functions, and one- and two-forms. The general rules of
integration can be got from what follows 
below and by linearity:
\begin{itemize}
\item[i)] For functions $z^{(\a, \b)}$,
\be
\VEV{z^{(\a, \b)}, (\a', \b' ) } = \d_{(\a, \a')} \d_{(\b, \b')}
:= \int_{(\a', \b')}
z^{(\a, \b)} = \VEV{ (\a', \b') | z^{(\a, \b)} | (\a', \b' ) }~. \label{4.8}
\ee
\item[ii)] For the one-form $\o^{(1)} = \o \otimes \unit$, where $\o$ is a
one-form for the first $L_N$, we have
\be
\VEV{\o^{(1)} , ( (\a, \b), (\a+1, \b )) } :=
\int_{(\a, \b)}^{(\a+1,\b)} \o^{(1)} =
\VEV{ (\a, \b)\, | \o^{(1)} F_+^{(1)\dag}  | (\a, \b ) }~,
\label{4.9}
\ee
$( (\a, \b), (\a+1, \b )) $ being the
link from $(\a,\b)$ to $(\a+1,\b)$.
If, more generally, the one--form is $\o^{(0)}\o^{(1)}$, where $\o^{(0)}$
is a generic function,
\bea
\int_{(\a, \b)}^{(\a+1,\b)} \o^{(0)} \o^{(1)} &=&
\VEV{(\a, \b)\, | \o^{(0)} \o^{(1)} F_+^{(1)\dag}  | (\a, \b )} \nonumber\\
&=&
\o^{(0)} [(\a,\b)] \VEV{(\a, \b) | \o^{(1)} F_+^{(1)\dag}  | (\a, \b )}~,
\label{4.10} \\
\o^{(0)}[\a,\b] &\equiv& \VEV{\o^{(0)},(\a,\b)}~.
\eea
Similarly, if $\wt{\o}^{(1)} = \g \otimes \o $ is a one-form and $\o^{(0)}$ a
function,
\bea
\int_{(\a, \b)}^{(\a,\b+1)} \o^{(0)} \wt{\o}^{(1)} &=&
\VEV{ (\a, \b)\, | \o^{(0)} \wt{\o}^{(1)} F_+^{(2)\dag}  | (\a, \b )}
\nonumber\\&=&
\o^{(0)} [(\a,\b)]\VEV{ (\a, \b)\, |
\wt{\o}^{(1)} F_+^{(2)\dag}  | (\a, \b ) }~. \label{4.11}
\eea
The integral of $ \o^{(1)}$ over $( (\a, \b), (\a, \b+1 )) $ and of $
\wt{\o}^{(1)}$ over $( (\a, \b), (\a+1, \b )) $ are zero.

\item[iii)] Next we 
come to two--forms. It is enough to consider the
integral of a two--form $\o^{(2)}$ over an elementary cell ${\cal C} =
((\a,\b),(\a+1,\b)) \times ((\a,\b),(\a,\b+1))$. We
define
\be
\int_{{\cal C}}\o^{(2)} = 
\VEV{(\a,\b)|\o^{(2)}{F_+^{(1)}}^\dagger
{F_+^{(2)}}^\dagger|(\a,\b)}. 
\ee

\item[iv)] Integrals of $k$-forms over $m$-cells are (of course) zero if
$k\not=m$.

\end{itemize}

It is important that there are analogues of Stokes' theorems valid in
this formulation. Consider first the integral of an exact one--form
$d\o^{(0)}$ over a series of touching links. Applying (\ref{4.6}) to
each link successively, we find Stokes' theorem for one-forms.

A corollary of this result is that the integral of an exact form over
a closed loop of links is zero. It means also that its
integrals from $A$ to $B$ by any two paths are equal, that is,
``continuous deformation of paths keeping ends fixed" does not change
the integral.

Finally, consider an exact two-form $d\o=F_+\o+\o F_+$, $\o$ being a
one--form. Its Stokes' theorem for the elementary cell $\cc$ would read
\be
\int_\cc d\o =\int_{\partial \cc}\o~.   \nonumber \\  
\ee

The proof is as follows. For the right-hand side, from (\ref{4.9}-\ref{4.11}) 
one gets 
\bea
\int_{\partial \cc}\o&= &\VEV{(\a,\b)|\o {F_+^{(2)}}^\dagger|(\a,\b)} + 
     \VEV{(\a,\b+1)| \o {F_+^{(1)}}^\dagger|(\a,\b+1)} \nonumber \\
&~& -\VEV{(\a,\b)|\o {F_+^{(1)}}^\dagger|(\a,\b)} - 
\VEV{(\a+1,\b)|\o {F_+^{(2)}}^\dagger|(\a+1,\b)}~. \label{4.13}
\eea

Now, $\o$ has the general form ~$\o = \xi f_+\otimes \unit + \eta \gamma 
\otimes
f_+$, where $\xi ,\eta $ are functions, while, from (\ref{3.22}), $F_+^{(1)} =
f_+
\otimes \unit~,~F_+^{(2)} = \g \otimes f_+$.  
In view of this, we can write (\ref{4.13}) as
\bea
\int_{\partial \cc}\o &= & \VEV{ (\a,\b) |~\h~|(\a,\b)} +
\VEV{(\a,\b+1) | ~ \xi ~|(\a,\b+1)} \nonumber
\\  &-& \VEV{(\a,\b) | ~ \xi ~|(\a,\b)} - \VEV{ (\a+1,\b) |~\h~|(\a+1,\b)} ~.
\label{sto1}
\eea

For the left-hand side, by writing $\xi = \sum_j \xi^{(1)}_j \otimes 
\xi^{(2)}_j$ and $\h = \sum_j \h^{(1)}_j \otimes \h^{(2)}_j$, one finds that
\bea
\int_\cc d \o &=& 
\VEV{(\a,\b)| d \o~{F_+^{(1)}}^\dagger {F_+^{(2)}}^\dagger |(\a,\b)} 
\nonumber \\
&=&\VEV{(\a,\b)| (F_+\o+\o F_+) {F_+^{(1)}}^\dagger {F_+^{(2)}}^\dagger
|(\a,\b)} \nonumber \\
&=& \bra{(\a,\b)} \left\{ - \sum_j f_+ \h^{(1)}_j {f_+}^\dagger 
\otimes \h^{(2)}_j
+ \sum_j \xi^{(1)}_j \otimes f_+\xi^{(2)}_j {f_+}^\dagger \right. \nonumber \\  
&~& \left. + \sum_j \h^{(1)}_j \otimes \h^{(2)}_j  
- \sum_j \xi^{(1)}_j \otimes \xi^{(2)}_j~ \right\} \ket{(\a,\b)}~. 
\label{1sto} 
\eea
Since $f_+ = 
{\scriptstyle            
 \addtolength{\arraycolsep}{-.5\arraycolsep}
 \renewcommand{\arraystretch}{0.5}
 \left( \begin{array}{cc}
 \scriptstyle 0 & \scriptstyle \Delta \\
 \scriptstyle 0  & \scriptstyle 0 \end{array} \scriptstyle\right)}$, 
on using (\ref{rotaz}), this becomes
\bea
\int_\cc d\o & = & \VEV{(\a,\b)|d \o {F_+^{(1)}}^\dagger {F_+^{(2)}}^\dagger
|(\a,\b)}
\nonumber \\
&= & - \VEV{ (\a+1,\b) |~\h~|(\a+1,\b)} + \VEV{(\a,\b+1) | ~ \xi
~|(\a,\b+1)} \nonumber \\  
&~& +\VEV{ (\a,\b) |~\h~|(\a,\b)} - \VEV{(\a,\b) | ~
\xi ~|(\a,\b)} ~, \label{2sto}
\eea
which is seen to coincide with (\ref{sto1}). Hence Stokes' theorem 
for $\cc$ is true.

In obtaining (\ref{1sto}-\ref{2sto}), we have for instance used the fact 
that $F_+^{(1)} {F_+^{(1)}}^{\dagger} \ket{(\a,\b)} =
\ket{(\a,\b)}$ since $\ket{\a}$ and $\ket{\b}$  are column vectors with zero
along their last $N$ rows.

 Stokes' theorem for a general cube follows by decomposing it into
elementary cells like $\cc$ and applying (\ref{4.13}) to each cell.

We can evidently extend these considerations to tori $T^N$ of arbitrary
dimensions.

\sxn{Gauge Theories}
\subsxn{The $*$-Algebra }

It is useful to change the algebra $\o^* \ca$ somewhat for physical
applications. The reason is that $\o^*\ca$ is not a 
\cstar  and so
causes (minor) problems in formulating reality (hermiticity)
conditions. Let us first attend to these modifications. 

We begin with the circle chains and its algebra (\ref{3.8}). We enlarge this
algebra by introducing new elements 
$\wt{z}^{k, k+1}$ with the properties
\be
z^{j, j+1} \sqcup \wt{z}^{k, k+1} = \wt{z}^{j, j+1}
\sqcup \wt{z}^{k, k+1} = 0~.
\label{5.1}
\ee

They are to be thought of as dual to  $z_{k,k+1}$, just as $z^{k,
k+1}$. But whereas in the definition of $z^j  \sqcup z^{k, k+1}$, 
the vertices had the order $[0, 1, 2, \dots, N-1]$, we use the
order $[N-1,N-2, \dots, 0]$ in the definition of $z^j 
\sqcup \wt{z}^{k, k+1}$,  so that 
\bea
&&z^j  \sqcup \wt{z}^{k, k+1} = \wt{z}^{k, k+1} \d_{j, k+1}~,
\nonumber \\
&& \wt{z}^{k, k+1} \sqcup z^j = \d_{jk} \wt{z}^{k, k+1}~.
\label{5.2}
\eea

Representing two elements 
$\sum \a_j z^j + \sum \a_{j, j+1} z^{j, j+1} + 
\sum \wt{\a}_{j, j+1} \wt{z}^{j, j+1}$ and   
$\sum \b_j z^j + \sum \b_{j, j+1} z^{j, j+1} + 
\sum \wt{\b}_{j, j+1} \wt{z}^{j, j+1}$ 
in terms of their components as 
\bea
&& \a = (\a_{0}, \a_{1}, \dots, \a_{N-1}; \a_{01}, \a_{12}, \dots \a_{N-1, N};
\wt{\a}_{01}, \wt{\a}_{12}, \dots \wt{\a}_{N-1, N})~, \nonumber \\
&& \b = (\b_{0}, \b_{1}, \dots, \b_{N-1}; \b_{01}, \b_{12}, \dots \b_{N-1, N};
\wt{\b}_{01}, \wt{\b}_{12}, \dots \wt{\b}_{N-1, N})~,  \label{5.3}
\eea
we now have for their product,
\bea
&& \a \b = (\a_0 \b_0, \a_1 \b_1, \cdots , \a_{N-1} \b_{N-1} ; 
\nonumber \\
&& ~~~~~~~~ \a_0 \b_{0,1} + \a_{0,1} \b_1,  \a_1 \b_{1,2} + \a_{1,2} 
		\b_2, \cdots,  
       \a_{N-1} \b_{N-1,N} + \a_{N-1,N} \b_N ; \nonumber \\
&& ~~~~~~~~ \a_1 \wt{\b}_{0,1} + \wt{\a}_{0,1} \b_0,  \a_2 \wt{\b}_{1,2} 
        + \wt{\a}_{1,2} \b_1, \cdots,  
       \a_{N} \wt{\b}_{N-1,N} + \wt{\a}_{N-1,N} \b_{N-1})~. \label{5.4} 
\eea

The algebra with product (\ref{5.4}) is a $^*$ (but not a $C^*$-)
algebra, the $^*$ operation being:
\be
\a^* = (\a_{0}^*, \a_{1}^*, \dots, \a_{N-1}^*; 
 \wt{\a}_{0, 1}^*, \wt{\a}_{1,2}^*, \dots \wt{\a}_{N-1,N}^*; 
 \a_{0,1}^*, \a_{1,2}^*,\dots, \a_{N-1,N}^*)~,
\label{5.5}
\ee
with the $*$ on the right hand side denoting the usual 
complex conjugation of complex numbers.
This
$*$ operation on the algebra fulfills $(\a\b)^* = \b^* \a^*$. 
We call this algebra as $\O^*\ca$. 
It may be worth noticing that the reconstruction theorem of 
Section~\ref{se:rt} works also for the algebra $\O^*\ca$ if one considers 
indecomposable differential ideals which are in addition $^*$ closed.

A convenient faithful representation of this algebra is as follows.
The functions 
$\ca := \O^{(0)} \ca$ are diagonal  $2N \times 2N$
matrices
\be 
a=\mbox{diag}[\a_{0},\dots ,  
\a_{N-1}; \a_{0}, \dots,  \a_{N-1}]~. \label{5.6}
\ee

As for one-forms $\O^{(1)} \ca$, they can be got from 
\be
F = \left( 
\begin{array}{cc}
0 & \D \\
\D^{\dag} & 0
\end{array}
\right) c~, \label{5.7}
\ee
$c$ being a fermionic annihilation operator. The role of $F$ now is
the previous role of $F_+$. The one-forms are thus $\sum a_i [F,
b_i]~, a_i, b_i \in \O^{(0)} \ca$. 

Two-forms are zero, $F^2$ being zero, the operator $c$ having been inserted for
this purpose. 

The $^*$ operation in this representation is hermitian conjugation on all
operators except $c$ :
\bea
&& a^* = a^{\dag}~~{\rm for}~~ a \in \O^{(0)} \ca~, \nonumber \\
&& (\sum a_i [F, b_i])^* = 
- \sum \left[
\left(
\begin{array}{cc}
0 & \D \\
\D^{\dag} & 0
\end{array}
\right), 
b_i^{\dag}\right]~a^\dagger _ic~.  \label{5.8}
\eea

Higher dimensional generalizations are easy. For $T^2$, we introduce two
mutually anticommuting fermion annihilation operators $c_1$ and $c_2$ and set 
\bea
F_1 &=& \left[ \left( 
\begin{array}{cc}
0 & \D \\
\D^{\dag} & 0
\end{array}
\right) \otimes \unit \right] c_1\nonumber\\
F_2 &=& \left[ \unit \otimes \left( 
\begin{array}{cc}
0 & \D \\
\D^{\dag} & 0
\end{array}
\right) \right] c_2~,  \nonumber \\
F &=& F_1 + F_2~.
\label{5.9}
\eea
Then the set of functions for the $T^2$-lattice is $\ca \otimes_{\IC} \ca$,
one-forms are $\sum f_i [F, g_i]$ and two-forms are $\sum f_i [F, g_i] 
[F, h_i]~, (f_i, g_i, h_i \in \ca \otimes_{\IC} \ca)$. There are no forms of
higher order. The $d$ operator on a form $\o^{(k)}$ of degree $k$ can be
expressed using $F$: 
\be
d \o^{(k)} = F \o^{(k)} - (-1)^k \o^{(k)} F~. \label{5.10}
\ee

The integration theory here is similar to that for $\o^*\ca$. It is
enough to replace $F_{+}^{(i) \dag}$ by $f_{+}^{(i) \dag} c_i^{\dag} $
in  the appropriate formulas of Section 4. Here,
$f_{+}^{(1)} = f_+ \otimes \II$ and $f_{+}^{(2)} = \II \otimes f_+$ with 
$f_+ = 
{\scriptstyle            
 \addtolength{\arraycolsep}{-.5\arraycolsep}
 \renewcommand{\arraystretch}{0.5}
 \left( \begin{array}{cc}
 \scriptstyle 0 & \scriptstyle \Delta \\
 \scriptstyle 0  & \scriptstyle 0 \end{array} \scriptstyle\right)}$. 
[Furthermore, of course traces such as that in (\ref{4.4}) now involve
also traces over creation-annihilation operators.]
Note that if $F_{+}^{(i) \dag}$ are now replaced by  $f_{+}^{(i)}c_i^{\dag}$
instead, we will get answers appropriate for integrals with orientations opposite
to the previous ones. 

For what follows, it is convenient to always call functions as $\ca :=
\O^{(0)}\ca$, forms of degree $k$ as $\O^{(k)}\ca$ and set 
$\O^* \ca =\oplus_k \O^{(k)} \ca $. This means that we rechristen 
$\ca \otimes_{\IC} \ca$
above as $\ca$. We also will not distinguish $\O^{(k)}\ca$ and $\O^*\ca$
from their representations.

\subsxn{The Yang-Mills Action}

Consider the `trivial' $\ca$-module (the algebraic counterpart of the 
module of sections of a trivial bundle)
\be
\ce = \IC ^N \otimes_{\IC} \ca := \ca^N~. \label{5.11}
\ee
Any element of $\ce$ will be written as a finite linear combination of `words'
\be
v \otimes a~, ~~v = (v_1, v_2, \dots, v_N)~, ~v_i \in \IC~, ~~ a \in \ca~.
\ee 
It is a right $\ca$-module, with the $\ca$ action simply given by
$ (v \otimes a) b = v \otimes ab$. We also introduce the spaces 
\bea
&& \ce^{(k)} := \ce \otimes_{\ca} \O^{(k)} \ca = 
\IC ^N \otimes_{\IC} \O^{(k)} \ca ~, 
\nonumber \\  
&& \ce^* = \oplus_k \ce^{(k)}~. \label{5.12}
\eea
Elements of ${\cal E}^{(k)}$ are (finite) combinations of terms like
$v^{(k)} = v \otimes_{\IC} 
\o^{(k)}$, with $v \in \IC^N$ and $\o^{(k)} \in \Omega^{(k)} {\cal A}$. Any 
${\cal E}^{(k)}$ is a right ${\cal A}$-module while $\ce ^*$ is a 
right $\O^*\ca $-module.

A connection on the module $\ce$ is a linear map 
$\nabla : \ce ^{(k)} \rightarrow \ce ^{(k+1)}$ which satisfies 
`Leibnitz rule'
\be
\nabla (v^{(k)} \r) = (\nabla v^{(k)}) \r + (-1)^k v^{(k)} d \r~, ~~ 
\forall~ v^{(k)} \in \ce ^{(k)}~, ~\r \in \O^*\ca ~. \label{leib}
\ee

A gauge potential can be introduced by defining
\be
\nabla = \II \otimes d + T(\a) \otimes A(\a) ~, \label{gapo}
\ee
summation on $\a$ being understood. The $A(\a)$'s are the connection one-forms
and the $T(\a)$'s are matrices  (they are taken to be a basis for some Lie
algebra, for instance a basis for the Lie algebra ${\underline U(N)}$ of $U(N)$;
see later).  It is immediate to check that $\nabla$ in (\ref{gapo}) satisfies 
(\ref{leib}). Explicitly,
\be
\nabla (v \otimes \o^{(k)}) = v \otimes d \o^{(k)} + T(\a) v \otimes A(\a) 
\o^{(k)}~.
\ee

By the very definition of $\nabla$, one checks that its square is $\O^* \ca 
$-linear, 
\be
\nabla^2 (v^{(k)} \r) = (\nabla^2 v^{(k)}) \r ~, ~~ 
v^{(k)} \in \ce ^{(k)}~, \r \in \O^*\ca ~. 
\ee
$\nabla^2$ is completely determinated by its restriction to $\ce$. Such a 
restriction defines the curvature $\cf(A)$ of the connection. One finds
\be
\cf(A)( v \otimes a) := \nabla^2 (v \otimes a) = 
\left( T(\a) v \otimes d A(\a) + T(\a) T(\b) v \otimes A(\a) A(\b) 
\right) a~.
\ee
We may therefore write
\be
\cf(A) = T(\a) \otimes d A(\a) + T(\a) T(\b) \otimes A(\a) A(\b)~. 
\label{curv} 
\ee

Bianchi identity now reads
\be
[\nabla, \cf(A)] = 0~. \label{bian}
\ee
This is really a trivial statement because left hand side is just 
$\nabla^3 - \nabla^3$. One can explicitly check it by using expressions 
(\ref{gapo}) and (\ref{curv}).

\bigskip

On $\ce^*$, there is an $\O^* \ca$-valued Hermitian structure given as follows.
Any two $\ce^{(k)}, \ce^{(k')}$ are orthogonal if $k \not= k'$,  while
\be
< v_1 \otimes \o_1^{(k)}, v_2 \otimes \o_2^{(k)} >_{k} ~:= 
< v_1 , v_2 > ~(\o_1^{(k)})^* \o_2^{(k)} := \sum_{j=1}^N 
v_{1j}^* v_{2j} ~(\o_1^{(k)})^* \o_2^{(k)}~. \label{herm}
\ee

Now, we have
\be
(d \o^{(k)})^* = (-1)^{(k+1)} d \o^{(k) *}~, 
\ee
so that compatibility of the connection with the Hermitian structure reads
\be
-< \nabla (v_1 \otimes a) , v_2 \otimes b > + 
<v_1 \otimes a , \nabla (v_2 \otimes b)> = d    
<v_1 \otimes a , v_2 \otimes b >~, a, b \in \ca~. \label{comp}
\ee
By using (\ref{gapo}) it reduces to
\be
-<T(\a) v_1, v_2>a^* A(\a)^* b ~+ <v_1, T(\a) v_2>a^* A(\a) b = 0~. \label{comp1}
\ee 
If the $T(\alpha)$'s are supposed to be hermitean, condition 
(\ref{comp1}) requires that the $A(\alpha)$'s  are ``real",
\be
A(\alpha)^*=A(\alpha)~. \label{real}
\ee

The continuum gauge group for $U_N$ now becomes its discretised version
${\cal U}_N$. It consists of $N \times N$ unitary matrices $g$ with $g_{ij}
\in {\cal A}$ \cite{CONLOT,CONNESBOOK,VARI}. It acts on ${\cal E}^{(k)}$
according to $e^{(k)} \rightarrow ge^{(k)}$ and transforms the connection
and curvature in the usual way:
\be
\nabla \rightarrow g \nabla g^{-1}~,~~{\cal F}(A)\rightarrow g {\cal F}(A)
g^{-1}~.\label{5.24}
\ee
Here the matrices $g$ are understood to be tensored with the appropriate unit
matrix. Furthermore, note that when dealing with representations, it is to be 
understood that functions like $g_{ij}$ are really of the form  
${\scriptstyle            
 \addtolength{\arraycolsep}{-.5\arraycolsep}
 \renewcommand{\arraystretch}{0.5}
 \left[ \begin{array}{cc}
 \scriptstyle g_{ij} & \scriptstyle 0 \\
 \scriptstyle 0  & \scriptstyle g_{ij} \end{array} \scriptstyle\right]}$
so that $g$ is to be thought of as
${\scriptstyle            
 \addtolength{\arraycolsep}{-.5\arraycolsep}
 \renewcommand{\arraystretch}{0.5}
 \left[ \begin{array}{cc}
 \scriptstyle g & \scriptstyle 0 \\
 \scriptstyle 0  & \scriptstyle g \end{array} \scriptstyle\right]}$.

Consider $\cf(A)^{\dagger}$
where $\dagger$ hermitean conjugates also
$c_i$. $\cf(A)^{\dagger}$ transforms in the 
same way as $\cf(A)$. 
Hence the discrete Euclidean Yang--Mills action in dimension $d$ can 
be taken to be
\be
S_{YM}(A)=\frac{1}{4g^2a^{4-d}}\Tr~ {\cal F}(A)^{\dagger}{\cal F}(A)~,
\label{5.26}
\ee
with $a$ being the lattice spacing and $g$ the coupling constant.
This expression is gauge invariant. Here Tr indicates a trace over 
all indices including a Hilbert space trace for the 
creation-annihilation operators.

For a gauge group other than ${\cal U}_N$, we must suitably further
restrict $g$ in the preceding discussion, and take $T(\alpha)$ to be the
generators of this group.

It is remarkable that (\ref{5.26}) is nothing but the Wilson action for a 
class of connections.
[For an alternative derivation of Wilson action for gauge fields based on
ideas of noncommutative geometry, see \cite{DMS}.]

Notice first that $F$ is a one-form
and therefore can be used as gauge potential.

The gauge potential $-F$ is a fixed point of the gauge group which we now take
to be $\cu_N$ for specificity :
\be
g(-F)g^{-1}+gdg^{-1}=g(-F)g^{-1}+g[F,g^{-1}]=-F~.\label{5.27}
\ee
So let 
\be
A=-F+\Phi ~.  \label{5.28}
\ee
Then $\Phi$ transforms homogeneously:
\be
g:~\Phi\rightarrow g \Phi g^{-1}~.\label{5.29}
\ee

In one dimension, suppose we write
\be
\Phi=\left(\begin{array}{cc}
0 & u\Delta \\
\Delta^{\dagger}u^{\dagger} & 0\\
\end{array}\right) c~, ~~u\in {\cal U}_N~, \label{5.30}
\ee
This expression is compatible with the reality condition on $A$. Also since 
$\Delta g^{-1}\Delta^{\dagger}\in {\cal U}_N$, it is also consistent with 
(\ref{5.24}), $u$ transforming according to
\be
u \rightarrow gu(\Delta g^{-1} \Delta^{\dagger})~~.\label{5.32}
\ee
This means that
\be
u(i) \rightarrow g(i)u(i)g(i+1)^{-1}  \label{5.33}
\ee
where the argument is the site index. [We are shifting 
it from subscript to argument, as the former will be assigned to symmetry 
indices]. But this is exactly the transformation law of the Wilson link
variable. Hence $u(i)$ can be identified with the link variable on link 
$z_{i,i+1}$ and $u(i)^\dagger$ with the one on $z_{i+1,i}$. 
Hereafter we will therefore often write
\be
u(i)=u_{i,i+1} \label{notation}
\ee
with similar notations in higher dimensions.

Now consider $T^2$. For $T^2$, the ansatz for $\Phi$, compatible with  
``reality" and stable under gauge transformations is
\bea
\Phi & = &
[\cu^{(1)} f_+^{(1)} + {f_+^{(1)}}^\dagger {\cu^{(1)}}^\dagger]c_1+
[\cu^{(2)} f_+^{(2)} + {f_+^{(2)}}^\dagger {\cu^{(2)}}^\dagger]c_2 \nonumber \\
& = & \xi c_1 + \eta c_2
\label{5.34}
\eea
where $\cu^{(k)}$ is really
${\scriptstyle            
\addtolength{\arraycolsep}{-.5\arraycolsep}
\renewcommand{\arraystretch}{0.5}
\left( \begin{array}{cc}
\scriptstyle u^{(k)} & \scriptstyle 0 \\
\scriptstyle 0  & \scriptstyle u^{(k)} \end{array} \scriptstyle\right)}$
and $(u^{(k)\dagger}u^{(k)})_{ij}=\delta_{ij}{\bf 1}$,
${\bf 1}$ being the constant
function in ${\cal A}$. The matrix $u^{(k)}(M)$ at site $M$ once more
transforms as a link in the direction $k$. We must thus think of
$u^{(k)}(M)$ as the Wilson variable on link from $M$ in the $k$-th
direction.

Generalizations to all dimensions must now be obvious.

Having obtained the Wilson link in our way, we can evaluate the action. 
Note first that
\be
{\cal F}(A)=\Phi^2~. \label{5.35}
\ee
Hence
\be
S_{YM}={1\over 4g^2a^{4-d}}\Tr (\Phi^{\dagger})^2\Phi^2~.\label{5.36}
\ee
{\em But this is just the Wilson action}.

As an example, consider the two-dimensional torus $T^2$. For $T^2$,
\be
\Phi^2=(\xi\eta-\eta\xi) c_1 c_2 \label{5.38}
\ee
so that we have
\be
S_{YM}=S^{(1)}_{YM}+ S^{(2)}_{YM}~~.\label{5.42}
\ee
where
\be
S^{(1)}_{YM}=-{1\over 2g^2a^2}\Tr \xi\eta\xi\eta ~, \label{5.40}
\ee
\be
S^{(2)}_{YM}={1\over 2g^2a^2}\Tr \xi^2\eta^2~. \label{5.41}
\ee

For each $(\alpha, \beta)$ in the notation of (\ref{4.8}), $S^{(1)}_{YM}$ 
has a term
\be
S^{(1)}_{YM}(\alpha \beta)= -{1\over g^2 a^2}\Tr \left\{
\Lambda_{cl}(\a,\b)+\Lambda_{acl}(\a,\b)
\right\} \label{5.43}
\ee
where $\Lambda_{cl}(\a,\b)$ is the product of the Wilson links around the
plaquette in the clockwise direction starting at $(\alpha, \beta)$, and
similarly $\Lambda_{acl}(\a,\b)$ for the anticlockwise direction.
A similar calculation gives
\be
S^{(2)}_{YM}(\alpha \beta)= {2 \over g^2 a^2} \Tr \II~. 
\ee
The contribution to the action of each $(\alpha, \beta)$ is therefore
\be
S_{YM}(\alpha \beta)={2\over g^2 a^2}\Tr \{ \II~ - {1 \over 2} \left(
\Lambda_{cl}(\a,\b)+\Lambda_{acl}(\a,\b)
 \right) \}~.
\ee
$S_{YM}$  is the sum over $\alpha, \beta$ of these terms and their
hermitean conjugates and is just the Wilson action. 

This calculation for $T^2$ readily generalizes to higher dimensions.

\sxn{The Dirac Action}
Let us consider a spacetime of dimension $K$. If $M$ is the
desired internal dimension of the Dirac field, introduce
\be
\wt{{\cal E}}= {\cal A} \otimes \IC ^M := {\cal A}^M = \left\{
e=(e_1, e_2, \cdots e_M); e_i \in {\cal A} \right\}~. \label{5.44}
\ee

Let us also consider
\be
 \tau_1 = \left(\begin{array}{cc} 0 & 1\\
1 & 0 \end{array} \right)
\label{tau}
\ee
acting on $\Omega^* \cal A$ and 
\be
T_1 = \tau_1 \otimes 1 \label{T}
\ee
acting on $\wt{\cal E}$.
     
Note that $T_1 $ is gauge invariant :
\be
\left(\begin{array}{cc}g & 0\\
0 & g \end{array}\right) T_1\left(\begin{array}{cc}g^{\dagger} & 0\\
0 & g^{\dagger} \end{array}\right) = T_1. 
\label{g}
\ee

The connection $\nabla$  acts on an element $e\in\wt{\ce}$ according to
\be
\nabla e =[ F+ A , T_1 ]e \label{5.45}.
\ee

Let us consider the connection $\nabla$ of the form (\ref{5.30}) and 
(\ref{5.34})  generalised to $K$ dimensions.
The component form of (\ref{5.45}) then is
\be
(\nabla e)_i =[ \Phi , T_1 ]_{ij}e_j, \qquad 1 \leq i,j \leq M,
 \label{component}
\ee
$i,j$ being internal indices.

We can clarify the meaning of (\ref{5.45}) further by examining the $K=1$
case. Then in the notation  where gauge transformation has the doubled form
$
{\scriptstyle            
 \addtolength{\arraycolsep}{-.5\arraycolsep}
 \renewcommand{\arraystretch}{0.5}
 \left( \begin{array}{cc}
 \scriptstyle g & \scriptstyle 0 \\
 \scriptstyle 0  & \scriptstyle g \end{array} \scriptstyle\right)}$, 
the right hand side of (\ref{component}) reads
\be
\left(\begin{array}{cc}(u_{ij}\D-\D^\dagger u^\dagger_{ij})e_j   & 0\\
0 & -(u_{ij}\D-\D^\dagger u^\dagger_{ij})e_j \end{array}\right)c
\label{array}
\ee
so that when $u=\II$, the upper and lower diagonals consist of forward 
and backward finite differences.

We next introduce the $\gamma$-matrices. Let $V$ be the vector 
space carrying the representation of $\gamma$'s \cite{Clifford}. Let
\be
{\cal E}_D = \wt{{\cal E}} \otimes _{\IC} V \ \ . \label{5.46}
\ee
An element of ${\cal E}_D$ is a ``spinor" $\psi$ with components
$\psi_{\lambda}^{\alpha}\in\ca ~~(\alpha=1,2,\cdots M;~\lambda=1,2,\cdots)$.

There are $K$ $\F^{(i)}$'s
corresponding to differentiations in $K$ directions. Therefore we can write
\be
\F=\sum \F^{(k)},
\label{A1}
\ee
$\F^{(k)}$ being proportional to $c_k$ in the $K-$dimensional version of 
(5.35).We can hence write
\be
(\nabla e)_i= \sum_k [\F^{(k)}, T_1]_{ij} e_j c_k
\equiv \sum D^{(k)}_{ij} e_j c_k.
\label{A2}
\ee

There being also $K$ $\gamma$-matrices $\gamma_k$, we can form the Dirac 
operator
\bea
D=\gamma_k D^{(k)}+m,\\ \nonumber
m= \mbox{Mass term}\ ,
\label{5.48}
\eea
and the Dirac equation
\be
[D \psi]^\a_\l \equiv (\g_k)^\r_\l D^{(k)}_{\a\b} \psi^\b_\r +m \psi^\a_\l=0.
\label{A3}
\ee
[There is no distinction between upper and lower internal indices, the internal
metric being the Kronecker delta.]

It is easy to check that (\ref{A3}) gives
Wilson's formulation \cite{LGT} for Dirac
equation in a doubled form . The source of the doubling is illustrated by
the appearance of {\it both } forward and backward differences in 
(\ref{array}) for $u=\II$. We get exactly Wilson's discretisation by 
retaining only one of these equations. 

We thus have a geometrical justification of the Wilson actions for gluons and
fermions which emerge naturally from the connection and curvature of a gauge
theory defined on a noncommutative lattice. It is however disappointing that
this approach is yet to suggest a solution for the doubling problem of chiral
fermions.

\sxn{Topological Action in 2d}

There is an analogue of the QCD $\theta$ term in the continuum 
two-dimensional QED. If $d A$ is the curvature, it is $\theta\! \int\! d A$.
Being independent of the metric, it is topological. It also integrates to an
integral of the connection, just as the $\theta$-term integrates to an
integral of the Chern-Simons term. The algebraic formulation and
noncommutative geometry have natural candidates for the discrete analogue
of this term, often called ``topological susceptibility". We now describe
these discretisations.

\subsxn{The First Choice}

The discretised curvature in this choice is ${\cal F}(A)=dA+A^2=\Phi^2$ 
[the $A^2$-term being present even for $U(1)$ gauge group.]. Hence the 2d 
lattice topological term is
\be
S_{\theta}= \theta \int(dA+A^2) + h.c. = \theta  
\Tr~\Phi^2F_+^{(1)\dagger}F_+^{(2)\dagger}+h.c. \label{6.1}
\ee
When $A$ is a pure gauge, ${\cal F}(A)=0$ and $S_{\theta}$ vanishes, as 
it should.

This proposal for topological susceptibility has a certain merit. It 
flows from the formalism without strain or doctoring, and also has the 
correct (naive) continuum limit.

But it is defective as well for the following reasons:

\begin{itemize}

\item[i)] It is not gauge invariant. Thus since $\Phi \rightarrow g \Phi 
g^{-1}$ under a gauge transformation $g$,
\be
\Tr~\Phi^2 F^{(1)\dagger}F^{(2)\dagger} \rightarrow 
\Tr~\left\{\Phi^2\left[g^{-1}F_+^{(1)\dagger}F_+^{(2)\dagger} g
\right] \right\} 
\label{6.2}
\ee
and $F_+^{(1)\dagger}F_+^{(2)\dagger}$ need not commute with $g$. The root of 
this difficulty is lack of graded commutativity in $\Omega^* {\cal A}$.

\item[ii)] It does not integrate to an integral of $A$.

\end{itemize}

Both these problems occur in 4d too. But it is possible to manage them 
in a reasonable manner in either dimension as we shall now see in 2d.

In 2d, as regards i), let us restrict gauge transformations by requiring that
\be
g^{-1} F_+^{(1)\dagger}F_+^{(2)\dagger}g=
F_+^{(1)\dagger}F_+^{(2)\dagger} ~. \label{6.3}
\ee

In the plaquette of Fig. 3, this means that the gauge transformations 
are constrained  to be the same at $(\alpha, \beta)$ and $(\alpha+1,
\beta+1)$. Thus in a square lattice with $N^2$ points, we can freely 
choose the value of $g$ at the $N$ points along say the bottom row.
This may not be too bad, this constraint  
superficially comparing favorably with the $1/N$ approximation 
\cite{Coleman}.
\begin{figure}[ht]
\begin{center}
\unitlength=1.00mm
\linethickness{0.4pt}
\begin{picture}(40.00,40.00)
\put(10.00,10.00){\circle*{1.33}}
\put(35.00,10.00){\circle*{1.33}}
\put(35.00,35.00){\circle*{1.33}}
\put(10.00,35.00){\circle*{1.33}}
\put(10.00,35.00){\line(1,0){25.00}}
\put(35.00,35.00){\line(0,-1){25.00}}
\put(35.00,10.00){\line(-1,0){25.00}}
\put(10.00,10.00){\line(0,1){25.00}}
\put(21.00,10.00){\vector(1,0){2.00}}
\put(22.00,35.00){\vector(1,0){1.00}}
\put(10.00,21.00){\vector(0,1){1.00}}
\put(35.00,21.00){\vector(0,1){1.00}}
\put(5.00,5.00){\makebox(0,0)[rt]{$(\alpha ,\beta)$}}
\put(40.00,5.00){\makebox(0,0)[lt]{$(\alpha +1,\beta)$}}
\put(40.00,40.00){\makebox(0,0)[lb]{$(\alpha +1, \beta +1)$}}
\put(5.00,40.00){\makebox(0,0)[rb]{$(\alpha ,\beta +1)$}}
\put(23.00,40.00){\makebox(0,0)[cb]{$W^-(\alpha ,\beta)$}}
\put(40.00,22.00){\makebox(0,0)[lc]{$W^+(\alpha ,\beta)$}}
\put(5.00,22.00){\makebox(0,0)[rc]{$W^-(\alpha ,\beta)$}}
\put(23.00,5.00){\makebox(0,0)[ct]{$W^+(\alpha ,\beta)$}}
\end{picture}
\end{center}
{\footnotesize {\bf Fig. 3.} Gauge invariance requires that gauge 
transformations  are the same at $(\a,\b)$ and $(\a+1,\b+1)$. 
$W^+(\a,\b)$ and $W^-(\a,\b)$ are the expressions in \eqn{6.5}.}
\end{figure}

The second limitation above  can be overcome if
\be
u^{(2)}F_+^{(1)\dagger}F_+^{(2)\dagger}u^{(2)\dagger}=
F_+^{(1)\dagger}F_+^{(2)\dagger}  \label{6.4}
\ee
or
$$
W^+(\alpha,\beta)=W^-(\alpha, \beta-1)  
$$
\be
W^-(\alpha, \beta):= tr~\left[u^{(2)}_{ad} u^{(1)}_{dc}
\right],~~~
W^+(\alpha, \beta):= tr~\left[u^{(1)}_{ab} u^{(2)}_{bc}\right]
\label{6.5}
\ee
where for notational simplicity, we have relabeled the points $(\alpha,
\beta),~(\alpha+1,\beta),~(\alpha+1,\beta+1),~(\alpha, \beta+1)$ as 
$a,b,c,d$.
Also we are using the notation \eqn{notation} so that, for example, 
$u^{(2)}[(\a\b)] = u^{(2)}_{(\a,\b)(\a,\b+1)}=u^{(2)}_{ad}$.
Note further that $\tr$ (with lower case  
$\tr$) is trace over internal indices. 
Although redundant in the present $U(1)$ case, we 
leave it in here for uniformity. (It will be removed in 
Section 7.2.)

This constraint is similar to the condition on $g$ above. 
It is fulfilled if
$u^{(2)}$ is invariant by translation from  $(\alpha,\beta)$ to 
$(\alpha+1, \beta+1)$. 
{\em It can be regarded as a gauge fixing condition, the constrained
gauge transformations (\ref{6.3}) identifying the left-over gauge 
group after the partial gauge fixing (\ref{6.4},\ref{6.5})}.
We will expand on this further in Section 8.

We have yet to show that $S_{\theta}$ integrates to a surface term with 
(\ref{6.4},\ref{6.5}). Now
\be
S_{\theta}=\theta\sum_{(\alpha,\beta)}\left[W^-(\alpha,\beta)-W^+(\alpha,\beta)
\right]  \label{6.6}
\ee

If $a^{\prime}[=(\alpha, \beta-1)]$ is the point vertically below $a$, then
$$
W^+(a)=W^-(a^{\prime})
$$
by (\ref{6.5}), and $W^-(a^{\prime})$ cancels $W^+(a)$
in the sum in \eqn{6.6}.

Suppose we are integrating over the rectangular lattice of Fig. 4. It 
follows from these cancellations that $S_{\theta}$ is a surface term: 
\be
S_{\theta}=\theta \sum_i [W^-(a^{\prime}_i)-W^+(b^{\prime}_i)]~.  \label{6.8}
\ee
\begin{figure}[ht]
\begin{center}
\unitlength=1.00mm
\linethickness{0.4pt}
\begin{picture}(90.99,100.00)
\put(10.00,15.00){\circle*{1.33}}
\put(25.00,15.00){\circle*{1.33}}
\put(40.00,15.00){\circle*{1.33}}
\put(10.00,30.00){\circle*{1.33}}
\put(25.00,30.00){\circle*{1.33}}
\put(40.00,30.00){\circle*{1.33}}
\put(10.00,50.00){\circle*{1.33}}
\put(25.00,50.00){\circle*{1.33}}
\put(40.00,50.00){\circle*{1.33}}
\put(55.00,50.00){\circle*{1.33}}
\put(75.00,50.00){\circle*{1.33}}
\put(90.00,50.00){\circle*{1.33}}
\put(10.00,65.00){\circle*{1.33}}
\put(25.00,65.00){\circle*{1.33}}
\put(40.00,65.00){\circle*{1.33}}
\put(55.00,65.00){\circle*{1.33}}
\put(75.00,65.00){\circle*{1.33}}
\put(90.00,65.00){\circle*{1.33}}
\put(10.33,85.00){\circle*{1.33}}
\put(25.33,85.00){\circle*{1.33}}
\put(10.33,100.00){\circle*{1.33}}
\put(25.33,100.00){\circle*{1.33}}
\put(55.00,15.00){\circle*{1.33}}
\put(75.00,15.00){\circle*{1.33}}
\put(90.00,15.00){\circle*{1.33}}
\put(55.00,30.00){\circle*{1.33}}
\put(90.00,30.00){\circle*{1.33}}
\put(40.33,85.00){\circle*{1.33}}
\put(75.33,85.00){\circle*{1.33}}
\put(90.33,85.00){\circle*{1.33}}
\put(40.33,100.00){\circle*{1.33}}
\put(75.33,100.00){\circle*{1.33}}
\put(90.33,100.00){\circle*{1.33}}
\put(10.00,15.00){\line(1,0){30.00}}
\put(90.00,15.00){\line(0,1){15.00}}
\put(40.00,15.00){\line(0,1){15.00}}
\put(25.00,30.00){\line(0,-1){15.00}}
\put(10.00,100.00){\line(1,0){30.00}}
\put(75.00,85.00){\line(0,1){15.00}}
\put(40.00,85.00){\line(0,1){15.00}}
\put(25.00,85.00){\line(0,1){15.00}}
\put(10.00,85.00){\line(0,1){15.00}}
\put(55.00,100.00){\circle*{1.33}}
\put(55.00,100.00){\line(-1,0){15.00}}
\put(75.00,100.00){\line(1,0){15.33}}
\put(40.00,15.00){\line(1,0){15.00}}
\put(55.00,15.00){\line(0,1){15.00}}
\put(75.00,15.00){\line(1,0){15.00}}
\put(17.00,15.00){\vector(1,0){1.00}}
\put(32.00,15.00){\vector(1,0){1.00}}
\put(46.00,15.00){\vector(1,0){1.00}}
\put(81.00,15.00){\vector(1,0){1.00}}
\put(17.00,100.00){\vector(1,0){1.00}}
\put(32.00,100.00){\vector(1,0){1.00}}
\put(46.00,100.00){\vector(1,0){1.00}}
\put(81.00,100.00){\vector(1,0){1.00}}
\put(25.00,21.00){\vector(0,1){1.00}}
\put(40.00,21.00){\vector(0,1){1.00}}
\put(55.00,21.00){\vector(0,1){1.00}}
\put(90.00,21.00){\vector(0,1){1.00}}
\put(25.00,91.00){\vector(0,1){1.00}}
\put(40.00,91.00){\vector(0,1){1.00}}
\put(10.00,91.00){\vector(0,1){1.00}}
\put(75.00,91.00){\vector(0,1){1.00}}
\put(21.00,30.00){\makebox(0,0)[rc]{$b_2$}}
\put(36.00,30.00){\makebox(0,0)[rc]{$b_3$}}
\put(86.00,30.00){\makebox(0,0)[rc]{$b_N$}}
\put(71.00,85.00){\makebox(0,0)[rc]{$a_n'$}}
\put(36.00,85.00){\makebox(0,0)[rc]{$a_3'$}}
\put(21.00,85.00){\makebox(0,0)[rc]{$a_2'$}}
\put(6.00,85.00){\makebox(0,0)[rc]{$a_1'$}}
\put(75.00,104.00){\makebox(0,0)[cb]{$a_n$}}
\put(40.00,104.00){\makebox(0,0)[cb]{$a_3$}}
\put(25.00,104.00){\makebox(0,0)[cb]{$a_2$}}
\put(10.00,104.00){\makebox(0,0)[cb]{$a_1$}}
\put(90.00,11.00){\makebox(0,0)[ct]{$b_n'$}}
\put(40.00,11.00){\makebox(0,0)[ct]{$b_3'$}}
\put(25.00,11.00){\makebox(0,0)[ct]{$b_2'$}}
\put(10.00,11.00){\makebox(0,0)[ct]{$b_1'$}}
\end{picture}
\end{center}
{\footnotesize {\bf Fig. 4.} Here is shown the rectangular lattice of 
integration. Only the surface terms indicated by arrows remain after
the gauge fixing condition \eqn{6.5}. }
\end{figure}

\subsection{Alternatives}

The boundary terms \eqn{6.8} suggest an alternative topological term which has
the virtue that it can dispense with the gauge fixing condition (\ref{6.5}).
But the constraint (\ref{6.3}) on the gauge group is still needed. 

The idea for this alternative follows from (\ref{6.8}) and the continuum 
results. We first write $
\sum_iW^-(a_i^{\prime})$ for example as a one-dimensional integral:
\be
\sum_iW^-(a_i^{\prime})=\theta 
\sum_{a_i} \VEV{a_i\left| \left\{\left( {f_+^{(2)}}^\dagger
u^{(2)} f_+^{(2)} \right)
\left(u^{(1)}{f_+^{(1)}}\right)\right\}{f_+^{(1)}}^\dagger
\right| a_i}~.  \label{6.9}
\ee
This is just the integral of the one-form
\be
{\cal A}=\left( {f_+^{(2)}}^\dagger\right) u^{(2)}
\left( {f_+^{(2)}}\right) 
\left( u^{(1)} f_+^{(1)}\right)c_1
\label{6.10}
\ee
over the top horizontal line. It is to be compared with the integral at the 
final time of the connection $A$ in the continuum theory.

Now the topological term in the latter is
$\theta \int dA$. In analogy, we can write its  
discrete analogue as follows:
\be
S^{\prime}_{\theta}=\theta \int d\ca~. \label{6.11}
\ee

Now with the constraint (\ref{6.3}) on $g$, $\cal A$ transforms homogeneously
and not as a connection under gauge transformations. For this (and other) 
reasons, $d {\cal A}$ is not gauge invariant. Still $S^{\prime}_{\theta}$ 
is, as it integrates to gauge invariant boundary terms. This is so 
without any condition on $u^{(j)}$. If we choose to impose the rule 
(\ref{6.4}) on $u^{(2)}$, then $S^{\prime}_{\theta}$ becomes $S_{\theta}$.

We can see this by just writing out $S^{\prime}_{\theta}$:
\newcommand{\due}{^{(2)}}
\newcommand{\uno}{^{(1)}}
\bea
S^{\prime}_{\theta}&=&\theta \sum_{(\a,\b)}
\left\langle 
(\a,\b)\left|f_+\due\left({f_+\due}^\dagger 
u\due f_+\due u\uno f_+\uno \right) {f_+\uno}^\dagger{f_+\due}^\dagger - 
\right.  \right.
\nonumber \\ 
& & - \left. \left.
\left({f_+\due}^\dagger u\due f_+\due u\uno f_+\due \right) 
{f_+\due}^\dagger f_+\uno {f_+\uno}^\dagger \right| 
(\a,\b)\right\rangle
\label{6.12}
\eea 
where the sum is over an appropriate range and the traces over
creation-annihilation operators have been performed. 
The first sum here gives the  $W^-(\a,\b)$ terms. On putting in
$u^{(2)}(\alpha,\beta)=u^{(2)}(\alpha+1,\beta+1)$, the second sum gives the 
$W^+(\a,\b)$ terms.

We can also begin from $W^+(b^{\prime}_i)$
to find another candidate 
$S^{\prime \prime}_{\theta}$ for topological susceptibility, giving us 
three possibilities in all for this term. They are all different if 
$u^{(2)}$ is not constrained, but become the same with a constrained 
$u^{(2)}$. Only future practice can decide the best choice among them or 
their combinations.

\sxn{ Topological Action in Four Dimensions}

We now turn our attention to the QCD topological term. In this case, 
the group being $SU(3)_c$, $\cal E$ is ${\cal A}^3$ and $u^{(i)}$ are 
both unitary and unimodular.

\subsxn{The First Choice}

There is as before an effortless construction of a discrete ``topological'' 
action ${\cal S}_{\theta}$ here as well. If ${\cal F}(A)$ is the 
curvature, it is 
$$
{\cal S}_{\theta}= {\theta \over 8 \pi^2} \int \tr~{\cal F}(A)^2 := 
{\theta \over 8 \pi^2} \Tr~{\cal F}(A)^2 \Pi_i f_+^{(i)\dagger}c_i^{\dagger}~,
$$
\be
\Pi_i f_+^{(i)\dagger}c_i^{\dagger} =
f^{(1)\dagger}_+ f^{(2)\dagger}_+
f^{(3)\dagger}_+ f^{(4)\dagger}_+
c_1^{\dagger }c_2^{\dagger }c_3^{\dagger}c_4^{\dagger}~.\label{7.1}
\ee
When $A$ is a pure gauge, ${\cal F}(A)=0$ and ${\cal S}_{\theta}$ 
vanishes, as it should.

It can be shown that the naive continuum limit of ${\cal S}_{\theta}$ is 
the continuum $\theta$-term.

Gauge invariance of ${\cal S}_{\theta}$ requires a constraint like 
(\ref{6.3}) on gauge transformations $g$, namely,
\be
g (\Pi_i f_+^{(i)\dagger}c_i^{\dagger}) = 
(\Pi_i f_+^{(i)\dagger}c_i^{\dagger}) g~.\label{7.2}
\ee
If the points of the $T^4$-lattice are labeled by 
$(\alpha,\beta,\gamma,\delta)$, 
(\ref{7.2}) means that
\be
g(\alpha+1,\beta+1,\gamma+1,\delta+1)=
g(\alpha,\beta,\gamma,\delta)~.\label{7.3}
\ee
Thus the value of $g$ along the diagonal is constrained to be a 
constant. In other words, if the number of points along each line of the
lattice is $N$ so that the total number of points is $N^4$, we can freely
choose $N^3$ out of $N^4$ values of $g$, the remaining ones being fixed by
(\ref{7.3}). Naively this appears an acceptable limitation.

Just as the 2d topological term, the QCD $\theta$-term in the continuum 
integrates to a surface term whereas like its two dimensional partner, ${\cal
S}_{\theta}$ does not do so without a gauge fixing condition. But before
stating this condition, let us write out ${\cal S}_{\theta}$. 

We define the hypercube $C(\alpha,\beta,\gamma,\delta)$ attached to the 
vertex $(\alpha,\beta,\gamma,\delta)$ as the cube with 
$(\alpha,\beta,\gamma,\delta)$ and $(\alpha+1,\beta+1,\gamma+1,\delta+1)$ 
as diagonally opposite vertices. Its vertices are 
$(\alpha +i, \beta +j, \gamma + k, \delta+l)$ where $i,j,k,l \in
\{0,1\}$. There are 4!=24 terms in ${\cal S}_{\theta}$ from each cube
$C(\alpha,\beta,\gamma,\delta)$ corresponding to that many permutations
of 1,2,3,4. Given one such permutation or order, say 3,1,2,4, we can pick 
a set of ordered vertices of the cube, those in this example being $\xi_0=
(\alpha,\beta,\gamma,\delta),~\xi_1=(\alpha,\beta,\gamma+1,\delta),~\xi_2=
(\alpha+1,\beta,\gamma+1,\delta),~\xi_3=
(\alpha+1,\beta+1,\gamma+1,\delta),~\xi_4=
(\alpha+1,\beta+1,\gamma+1,\delta+1)$. Thus for each permutation $P$, 
there are five ordered vertices from $(\alpha,\beta,\gamma,\delta)$
to the diagonally opposite corner $(\alpha+1,\beta+1,\gamma+1,\delta+1)$. 
Now join the successive vertices to get four links and take the trace of 
the product of the Wilson variables along these links. Multiplying this 
expression by ${\theta \over 8 \pi^2} (-1)^{\epsilon(P)}$, where $\epsilon(P)$
is the  signature of $P$, we get the contribution of $P$ to ${\cal S}_{\theta}$. 
In the example above, this contribution is
\be
 {\theta \over 8 \pi^2} ~\Tr~
u^{(3)}_{\xi_0 \xi_1} u^{(1)}_{\xi_1 \xi_2}
u^{(2)}_{\xi_2 \xi_3} u^{(4)}_{\xi_3 \xi_4}~~\label{7.4}
\ee

We can concisely write a typical term as ${\theta \over 8 \pi^2}
(-1)^{\varepsilon(P)}  W_{P(1234)}(\alpha\beta\gamma\delta)$, (\ref{7.4}) being 
${\theta \over 8 \pi^2} W_{3124}(\a\b\gamma\d)$. The 
total contribution ${\theta \over 8 \pi^2} W(\alpha\beta\gamma\delta)$ 
from $C(\alpha\beta\gamma\delta)$ is its sum over $P$,
\be
 {\theta \over 8 \pi^2} 
W(\alpha\beta\gamma\delta) =  {\theta \over 8 \pi^2}\sum _P
(-1)^{\varepsilon(P)} 
W_{P(1234)}(\alpha\beta\gamma\delta) 
\label{7.5}
\ee
while
\be
{\cal S}_\theta = 
 {\theta \over 8 \pi^2}
\sum_{(\alpha\beta\gamma\delta)} W(\alpha\beta\gamma\delta).
\label{7.6}
\ee

This expression seems a great deal simpler than 
existing proposals for topological susceptibility which typically contain
products of eight
$u^{(i)}$'s per hypercube \cite{Luscher,GKLSW,Teper} while there are only 4 here.

We have already addressed the question of gauge invariance of ${\cal
S}_\theta$. It remains to discuss the analogue of item {\em ii)} in $2d$.
Just as in $2d$, the continuum integral of 
$\tr F\wedge F$ integrates to a
surface integral. But ${\cal S}_\theta$ does so only with a gauge condition,
again like in $2d$, as we now explain. 

We can write
\bea
W(\alpha\beta\gamma\delta) &=& W^+(\alpha\beta\gamma\delta) -
W^-(\alpha\beta\gamma\delta) \label{7.7}\ \ ,\\
W^\pm {(\a\b\gamma\d)} & = & 
\sum_{\varepsilon(P)=\pm 1} W_{P(1234)}(\alpha\beta\gamma\delta)
\label{7.8}\ \ .
\eea

The analogue of the $2d$ gauge condition now is
\be
W^+\left(\alpha\beta\gamma\delta\right)=W^-(\alpha\beta\gamma(\delta-1))\ \ .
\label{7.9}
\ee
It cannot be written in a form like \eqn{6.4} because of the noncommutativity 
of the nonabelian $u^{(i)}$'s.

It should be evident that ${\cal S}_\theta$ integrates to a surface term 
with \eqn{7.9}. If the $\delta$--direction is thought of 
as time direction, and 
$\delta\in\{1,2,\ldots,N\}$, then ${\cal S}_\theta$ consists of integrals at 
future and past time boundaries:
\be
{\cal S}_\theta = 
 {\theta \over 8 \pi^2} \sum_{(\alpha\beta\gamma)}
\{  W^+(\alpha\beta\gamma1) - W^-(\alpha\beta\gamma(N-1)) \}~. \label{7.10}
\ee

In four dimensions, the analogue of 
$W(\alpha\beta\gamma\delta)$
consists of sums of terms, a typical one being 
like \eqn{7.4}, with 
$u^{(j)}_{\eta\xi}$ interpreted now as
$\mbox{P}\exp\int_\xi ^\eta A$. 

The expression \eqn{7.10} can be further expanded in lattice spacing,
the first nontrivial term then involves the connection. In a similar
manner the first nontrivial term in 
$W(\alpha\beta\gamma\delta)$
involves the Chern--Simons 3-form. The gauge condition \eqn{7.9} in
the continuum corresponds to a constraint on this form.

\sect{The Chern--Simons Term} \label{se:cs}

In the continuum the Chern--Simons (CS) term exists in all odd
dimensions, and is the integral of the CS form \cite{CS}. It has
played a crucial role in many recent studies in theoretical
physics. There have also been previous proposals for its discrete
analogue \cite{PhillipsStone,Semenoff}.

In this section we briefly show that there are candidates for this
term in the algebraic approach. For brevity we will limit ourselves to
1 and 3d. The candidates we will propose fail to have several of the
properties of the continuum Chern--Simons in an exact manner,
fulfilling them only up to corrections vanishing in the naive
continuum limit.

One candidate in the discrete for this term is obtained by integrating
the discretized form of CS density. It has the expression
\be
{k \over 8\pi^2} \int \tr(AdA+{2\over 3} A^3)= 
{k \over 8\pi^2} \Tr(AdA+\frac23 A^3) 
f_+^{(1)\dagger} 
f_+^{(2)\dagger} f_+^{(3)\dagger} c_1^{\dagger }c_2^{\dagger}c_3^{\dagger }
\label{8.1}
\ee
for any gauge group, where $k$ is a normalization constant.

Lack of graded commutativity means that $AdA\neq dA\,A$. For this
reason, it seems best to symmetrize $AdA$ and write for a discrete CS
term,
\bea
{\cal A}_{CS} & = & {k \over 8\pi^2} 
\Tr \left({1\over 2} (AdA + dA\,A)+{2\over
3}A^3\right) {f_+^{(1)}}^{\dagger }{f_+^{(2)}}^{\dagger }{f_+^{(3)}}^{\dagger }
c_1^{\dagger }c_2^{\dagger }c_3^{\dagger }\nonumber\\
& = & {k \over 8\pi^2} \Tr \left({1\over 2} (A{\cal F}(A) + {\cal F}(A)A)-{1\over
3}A^3\right) {f_+^{(1)}}^{\dagger }{f_+^{(2)}}^{\dagger }{f_+^{(3)}}^{\dagger }
c_1^{\dagger }c_2^{\dagger }c_3^{\dagger }.\label{8.2}
\eea
Substituting $A=-F+\Phi$ and ${\cal F}(A)=\Phi^2$, it becomes
\be
{\cal A}_{CS} = {k \over 8\pi^2} \Tr \left( {2\over 3}\Phi^3 - {1\over 6}(F\Phi^2+\Phi^2 F) +
{1\over 3}(\Phi F\Phi -F\Phi F)
\right) {f_+^{(1)}}^{\dagger }{f_+^{(2)}}^{\dagger }{f_+^{(3)}}^{\dagger }
c_1^{\dagger }c_2^{\dagger }c_3^{\dagger } ~. \label{8.3}
\ee

\sxn{The Hamiltonian Formulation}

In the Hamiltonian formulation, the poset is the topological lattice of the
spatial manifold. The constructions of the preceding sections, and especially 
of Section 5, are still largely
valid if physical interpretations are suitably modified and names are changed
accordingly. In particular, $dA + A^2$ is to be identified with the magnetic
field $B(A)$,
\be
B(A) = dA+A^2 := {\cal F}(A)
\label{9.1}
\ee
and $S_{YM}(A)$ with the ``potential energy" term (proportional to) $V(A)$ of the
Yang-Mills Hamiltonian:
\be
V(A) = \Tr B(A)^\dagger B(A)\ \ .
\label{9.2}
\ee

It remains to define the electric field, the Gauss law constraint and finally
the Hamiltonian. We take advantage of earlier work of Rajeev \cite{Rajeev} in 
formulating these definitions.

Let us introduce an orthonormal basis $\phi_n$ for the Lie algebra--valued 
one--forms, as it will be useful shortly:
\bea
(\phi_n, \phi_m ) &=& \Tr \phi_n^\dagger \phi_m = \delta_{nm}, \nonumber \\
\phi_n &=& \phi^\alpha_n\, T(\alpha), \qquad \phi^\alpha_n \in \Omega^1
\ca~,  ~(\phi^\alpha_n)^* = \phi^\alpha_n.
\label{9.3}
\eea

We will work in the Schr\"odinger representation where wave functions 
are  ${\com}$-valued functions of $A$.

In the continuum, the electric field $E= ( E_1,\, E_2, \cdots, E_M), \; M =
K-1, ~E_i = E^\alpha_i T(\alpha)$, consists of conjugate operators to $A_j =
A^\alpha_j T(\alpha)$ [$M$ being the spatial dimension]. If $\chi$ is a
functional of
$A$, then, at a formal level, $i E^\alpha_i(x)$ is the operator
${{\delta}\over{\delta A^\alpha_i (x)}}$, $x$ being a spatial point. Hence the
test function space for $E$ consists of Lie algebra-valued one forms, $f=
(f_1,\,f_2,\cdots,f_M), \; f_i = f^\alpha_i T(\alpha),
(f^\alpha_i)^*=f^\alpha_i$, the pairing $\VEV{f,E}$ acting on $\chi$ according to
\cite{Rajeev}
\be
\left( i \VEV{f,E} \chi \right) (A) = \lim_{t \ra 0} {{\chi( A+tf) -
\chi(A)}\over{t}}, \qquad t\in {\real}.
\label{9.4}
\ee

In a similar manner, for a poset, 
we introduce the pairing $\VEV{\lambda,E}$ for 
$\lambda \in
\underline{{\cal U}_N} \otimes (\Omega^1\ca)$, where
$\underline{{\cal U}_N} $ is the Lie algebra of ${\cal U}_N$. 
Thus $\lambda$ 
is a Lie algebra valued one-form, $\lambda = \lambda(\alpha) T(\alpha)$. We
also require that 
\be
\lambda(\alpha)^* = \lambda(\alpha)
\ee
just as $A(\alpha)^* = A(\alpha)$. 
 
$\VEV{\lambda,E}$ acts on wave functions according to
\be
\left( i \VEV{\lambda,E} \psi\right ) (A) = \lim_{t \ra 0} {{\psi(A + t\lambda) -
\psi(A)} \over {t}}. 
\label{9.5}
\ee

There is also an elegant statement of Gauss' law in this formulation. Let
\be
\Lambda \in \underline{{\cal U}_N} \otimes \ca
\label{9.6}
\ee
so that
\be
\Lambda = \Lambda(\alpha) T(\alpha), \qquad \Lambda(\alpha) \in \ca\ \ .
\label{9.7}
\ee
$\Lambda$ is a Lie algebra-valued function. Let us also require that
\be
\Lambda(\alpha)^* = \Lambda(\alpha)
\ee
The covariant derivative of $\Lambda$ is
\be
\nabla_A \Lambda \equiv [F,\Lambda^\alpha] T(\alpha) + A^\beta \Lambda
^\alpha 
[T(\beta), T(\alpha)].
\label{9.8}
\ee

The Gauss' law is just the condition
\be
\VEV{ \nabla_A \Lambda, E} \psi = 0
\label{9.9}
\ee
on the physical states $\psi$.

The Yang-Mills Hamiltonian is the operator
\be
H = {\frac {1}{2e^2}} \left[ \sum_n (\phi_n ,E)^\dagger(\phi_n,E) +
\hat{V}\right]
\label{9.10}
\ee
where
\be
(\hat{V} \psi)(A) = V(A) \psi(A)
\label{9.11}
\ee
and the first term in $H$ is to be defined using (\ref{9.5})
and the scalar product below.

Finally, the scalar product on wave functions is
\be
(\psi,\chi) = \int (\Pi_\alpha dA^\alpha ) \psi^* (A) \chi(A).
\label{9.12}
\ee

The Hamiltonian gauge theory on topological lattices is defined by this scalar
product, the Gauss' law (\ref{9.9}) and the Hamiltonian (\ref{9.10}). It
describes a bunch of oscillators with the exotic restriction (\ref{9.9}) on
wave functions.

There  is one more result we wish to discuss here, namely the use of
the CS functional in the Hamiltonian context. As is 
well known, with its help, we can readily change
the $\theta$-angle associated with a state vector $\psi$ for $M=3$ [Cf.\ 
Jackiw in \cite{CS}].

Thus let us suppose that a wave functional $\psi_0$ in the continuum is 
invariant under {\em all} gauge transformations, including those not
vanishing at infinity. Then $\psi_0$, being invariant under {\it all}
gauge transformations, corresponds to the continuum gauge theory with zero
$\theta$-angle. The wave functional $\psi_\theta$ in the continuum gauge
theory with non-zero $\theta$ is given by
\be
\psi_\theta (A) = \left[ \exp i \theta C(A) \right] \psi_0 (A)\ \ .
\label{9.13}
\ee
$C(A)$ being the continuum CS term.

In an analogous manner we can try using a CS term of Section \ref{se:cs}
to change the $\theta$ angle of a quantum state in the discrete context.
We have not however explored this possibility in any depth.

The discussion of this Section can be straightforwardly adapted also to 
groups such as $SU_N$.

\sxn{Final Remarks}

In this paper we have made progress on two fronts in formulating quantum 
physics on lattices. The first front is conceptual and concerns methods to 
approximate topology as well in the discretization of continua. In our 
approach to this important issue, we have adopted Sorkin's ideas 
\cite{SORK,SorkinGrav} to 
topologize cubical and simplicial decompositions of manifolds. We have then 
argued that the noncommutative algebra of cochains under the cup product fully 
captures the above topological data. In addition it is also perfectly adapted 
for quantum physics when combined with the Connes--Lott technology for 
noncommutative geometry \cite{CONLOT,CONNESBOOK,VARI}.

In this manner we have developed lattice gauge theories for cubical lattices 
and shown the natural emergence of Wilson's action therefrom 
for gauge and spin half fields.

The second front where this paper shows progress concerns topological actions 
like the QCD $\theta$ term and the Chern--Simons action. Our proposals for 
discretizations have several apparently superior features to alternative 
existing models.

There is one aspect regarding these topological terms which bears
emphasis. It seems impossible to formulate their discrete analogue without
losing, at least in an approximate way, one or another of their basic
properties. The root of this difficulty is the well--known impossibility
of converting cochains into an associative, graded commutative algebra.
This could be an indication that their reguralization in continuum
interacting field theories leads to anomalies. The latter may be
intolerable, for example they may disturb gauge invariance. If so, that
would be a good reason to choose $\theta=0$ for example in QCD, thereby
also resolving the strong CP problem \cite{CP}.

In this paper we have emphasized the action functional formulation. 
It is what is needed for functional integral quantization. Despite its 
superiority for numerical 
work over the Hamiltonian, the latter as well has its 
place in discrete physics. Therefore we have also devoted a section for its
treatment.

The substantial portion of this paper concentrates on cubical 
decompositions of 
the manifold. Preliminary explorations of the algebraic approach for 
simplicial decompositions have also been made with Allen Stern \cite{Stern}. A 
notable surprise we have encountered in this work is that it is not in general 
possible to realize the differential $d$ as a graded commutator with an 
operator like $F_+$ or $F$. We hope to report on this result 
elsewhere.

The effectiveness of our topological approach to discrete quantum physics and 
its corresponding algebraic description is not limited to gauge theories. 
It is very useful for 
example for soliton physics, and preserves fragile but important features like 
winding numbers, ruined by ordinary discretizations. This point has been
already explained elsewhere \cite{BBET} and will be thoroughly developed
in a forthcoming work.

There is one other setting in which these methods lead to interesting results. 
Preliminary work indicates that our discrete quantum physics may be effective 
for studying topology change and also for formulating discrete topological 
quantum field theories. We plan to report on this research after it meets a 
measure of success.

\ \\
{\large \bf Acknowledgments}\\
We acknowledge with gratitude the collaboration of Elisa Ercolessi, Arshad 
Momen and Gianni Sparano during the early stages of this work, and our 
fruitful discussions with them, Beppe Marmo, Peter Michor, 
Apoorva Patel and Al Stern
about its diverse aspects. In particular the relevance of cup algebra to 
our research was brought to our attention by Peter. Much of this work was 
carried out during our stay at the Erwin Schr\"odinger Institute for 
Mathematical Physics (ESI) in Vienna. We also thank ESI for financial 
support, and Beppe 
and Peter for making our stay in Vienna a rewarding 
experience. This work was in addition supported by the 
Istituto Italiano di Studi Filosofici
by the Department of Energy U.S.A.\ under 
contract numbers DE-FG-02-84ER40173 and, DE-FG02-85ER40231, and by INFN (Italy).

\vfill\eject

\end{document}